\documentclass[aps,prl,superscriptaddress,twocolumn,10pt]{revtex4-1}

\usepackage{graphicx,amsfonts,amsmath,amssymb,amstext}
\usepackage{float,wrapfig}
\usepackage{subfigure, psfrag}
\usepackage{dsfont}
\usepackage{color}
\usepackage{bm}

\newcommand{\be}{\begin{equation}}
\newcommand{\ee}{\end{equation}}
\newcommand{\ba}{\begin{eqnarray}}
\newcommand{\ea}{\end{eqnarray}}
\newcommand{\sign}[1]{\,\mbox{sgn}\left({#1}\right)}

\definecolor{purple}{rgb}{0.8,0,0.6}
\definecolor{darkgreen}{rgb}{0.00,0.6,0.00}

\def\mytitle{Odd-frequency Berezinskii superconductivity in Dirac semimetals}

\begin{document}

\title{\mytitle}
\date{August 29, 2019}

\author{P.~O.~Sukhachov}
\email{pavlo.sukhachov@su.se}
\affiliation{Nordita, KTH Royal Institute of Technology and Stockholm University, Roslagstullsbacken 23, SE-106 91 Stockholm, Sweden}

\author{Vladimir~Juri\v{c}i\'{c}}
\email{vladimir.juricic@nordita.org}
\affiliation{Nordita, KTH Royal Institute of Technology and Stockholm University, Roslagstullsbacken 23, SE-106 91 Stockholm, Sweden}

\author{A.~V.~Balatsky}
\email{avb@nordita.org}
\affiliation{Nordita, KTH Royal Institute of Technology and Stockholm University, Roslagstullsbacken 23, SE-106 91 Stockholm, Sweden}
\affiliation{Department of Physics, University of Connecticut, Storrs, CT 06269, USA}

\begin{abstract}
We formulate a general framework for addressing both odd- and even-frequency superconductivity in Dirac semimetals and demonstrate that the odd-frequency or the Berezinskii pairing can naturally appear in these materials because of the chirality degree of freedom. We show that repulsive frequency-dependent interactions favor the Berezinskii pairing while an attractive electron-electron interaction allows for the BCS pairing. In the case of compensated Dirac and Weyl semimetals, both the conventional BCS and odd-frequency Berezinskii pairings require critical coupling. Since these pairings could originate from physically different mechanisms, our findings pave the way for controlling the realization of the Berezinskii superconductivity in topological semimetals. We also present the density of states with several cusp-like features that can serve as an experimentally verifiable signature of the odd-frequency gap.
\end{abstract}

\maketitle

{\em Introduction.---}
The odd-frequency (OF) superconductivity, which was first suggested by Berezinskii in 1970s as a possible order parameter for superfluid He$^3$~\cite{Berezinskii:1974}, continues to be a challenging and interesting problem both from the theoretical and experimental perspective (for reviews of the OF superconductivity, see Refs.~\cite{Tanaka-Nagaosa:rev-2012,Linder-Balatsky:rev-2017,Triola-Black-Schaffer:rev-2019}).

Following the initial attempts with the phonon-mediated interactions in Refs.~\cite{Balatsky-Abrahams:1992,Abrahams-Allen:1993}, it was proposed that the spin-dependent fluctuations might lead to the realization of the Berezinskii pairing~\cite{Abrahams-Allen:1993,Fuseya-Miyake:2003}.
Later, the OF paring was considered in the Hubbard models for strong-coupling electron-phonon systems~\cite{Shigeta-Tanaka:2009,Kusunose-Miyake:2011b,Shigeta-Tanaka:2012}. However, the fundamental question about the microscopic mechanism of the OF superconductivity still awaits clarification. It is known that an OF pairing requires the electron-electron interaction with strong frequency-dependence. Compared with the infrared divergence governing an even-frequency (EF) pairing, an OF one seems to be disfavored in the bulk of conventional metals.

There are now numerous examples where the OF or Berezinskii superconducting states might naturally appear in various condensed matter systems including heterostructures~\cite{Tanaka-Golubov:2007a,Tanaka-Golubov:2007b,Eschrig-Schon:2007}, multiband~\cite{Black-Schaffer-Balatsky:2013,Triola-Balatsky:2017,Triola-Black-Schaffer:rev-2019} and driven~\cite{Triola-Balatsky:2016} systems, vortices in the type-II superconductors~\cite{Yokoyama-Golubov:2008}, to name but a few examples.

The key relation that guides the search of unconventional superconductors is the $SP^{*}O T^{*}=-1$ rule~\cite{Linder-Balatsky:rev-2017}, where $S$, $P^{*}$, $O$, and $T^{*}$ state for the spin, coordinate, orbital, and time permutations. This relation allows one to easily quantify the symmetry properties of the OF pairing. It also provides a guidance to predict possible pairing states in the multiband and chiral systems like Dirac and Weyl semimetals.

In this paper, we provide a general theoretical scheme that is able to study the OF superconductivity in nodal Dirac and Weyl systems.
First, by using the effective action approach~\cite{Solenov-Mozyrsky:2009,Kusunose-Miyake:2011}, we consider the possibility of the intrinsic OF pairing in Dirac semimetals ~\cite{Wehling-Balatsky:rev-2014,Yan-Felser:2017-Rev,Hasan-Huang:2017-Rev,Armitage-Vishwanath:2017-Rev}, which are natural platform for the topologically nontrivial superconductivity~\cite{Bernevig,Schnyder-rev:2015,Sato:2016evq}.
Next, we propose a microscopic scenario for realizing the OF Cooper pairs due to a \emph{repulsive} frequency-dependent interaction.
This possibility is plausible since the OF gap is determined by the derivative of the  with respect to frequency rather than the potential itself.
Therefore, the \emph{repulsive} pairing potential can lead to an \emph{effectively attractive} potential for the Berezinskii pairing. This allows for a completely different mechanism for a superconducting pairing, which is usually absent for the EF case. Finally, we demonstrate that the generation of both even- and odd-frequency gaps in compensated Dirac semimetals requires values of the potential strengths that exceed the critical ones. This provides the possibility to rule out the ubiquitous in conventional materials BCS type of superconductivity. As an experimental signature, the distinctive cusp-like features in the density of states (DOS) are identified.
We believe also that our work opens up new directions in studying superconductivity in dynamical systems. In particular, it is important in view of a possible realization of odd-frequency states in interacting organic Dirac materials reported in Ref.~\cite{Hirata-Kanoda:2017}
and in twisted bilayer graphene~\cite{Cao-Herrero:2018}.

{\em Model.---}
OF pairing is essentially a time-dependent pairing state, where the inclusion of the retardation effects is crucial to properly describe this phenomenon. Therefore, we choose not to use the effective Hamiltonian language that is challenging and debated~\cite{Solenov-Mozyrsky:2009,Kusunose-Miyake:2011,Fominov-Eschrig:2015}. Instead, we employ the effective action approach. The mean-field (MF) form of the effective action, which is valid for both even- and odd-frequency superconductors, reads (for the details of the derivation, see Sec.~II in the Supplemental Material (SM))
\begin{widetext}
\begin{eqnarray}
\label{model-OF-S-MF-1}
S_{MF} = -\frac{1}{2}\int dx_1 dx_2 \bar{\Psi}_{\rm N}(x_1) G^{-1}_{\rm N}(x_1-x_2) \Psi_{\rm N}(x_2) +\frac{1}{2} \int dx_1 dx_2 \frac{|\Delta_{MF}(x_1-x_2)|^2}{V(x_1-x_2)},
\end{eqnarray}
\end{widetext}
where $x=\left(t,\mathbf{r}\right)$ is the time-space four-vector, $\Delta_{MF}(x_1-x_2)$ is the mean-field gap, $V(x_1-x_2)$ corresponds to the interaction potential, $\Psi_{\rm N}(x)=\left\{\Psi(x),\hat{\Theta}\Psi(x)\right\}^{T}$ is the wave function in the Nambu space, $\hat{\Theta}= i \mathds{1}_2\otimes\sigma_y \hat{K}$ is the time-reversal (TR) operator, $\mathds{1}_2$ is the unit $2\times2$ matrix, $i\sigma_y $ is the Pauli matrix that describes the spin flip, and $\hat{K}$ is the complex conjugation operator. In addition, throughout this paper, we set $\hbar=c=k_B=1$. The inverse Green's function $G^{-1}_{\rm N}(x_1-x_2)$ is
\begin{eqnarray}
\label{model-OF-G-1-def}
&&G^{-1}_{\rm N}(x_1-x_2) = \left[i\partial_{t_1}  -\hat{H}_{\rm N}(x_1)\right] \delta(x_1-x_2) \nonumber\\
&&- \frac{\tau_x+i\tau_y}{2} \Delta_{MF}(x_1-x_2) - \frac{\tau_x-i\tau_y}{2} \Delta^{\dag}_{MF}(x_1-x_2).
\end{eqnarray}
Here $\bm{\tau}$ are the Pauli matrices acting in the Nambu space. The Nambu Hamiltonian $\hat{H}_{\rm N}(x)$ is defined as
\begin{equation}
\label{model-OF-HBdG}
\hat{H}_{\rm N}(x) = \frac{\mathds{1}_2+\tau_z}{2} \delta(t)\hat{H}(\mathbf{r}) -\frac{\mathds{1}_2-\tau_z}{2} \delta(t) \hat{\Theta} \hat{H}(\mathbf{r}) \hat{\Theta}^{-1},
\end{equation}
where $\hat{H}(\mathbf{r})$ for 3D Dirac semimetals (DSM) is defined in Eq.~(\ref{model-free-Hamiltonian-Weyl-rel}) below.

The mean-field gap in the momentum space reads
\begin{equation}
\label{Gap-Eq-def}
\hat{\Delta}(\omega, \mathbf{k}) = -i \int\frac{d \omega^{\prime} d^3k^{\prime}}{(2\pi)^4} \hat{V}\left(\omega-\omega^{\prime}; \mathbf{k}- \mathbf{k}^{\prime}\right) \hat{F}(\omega^{\prime}, \mathbf{k}^{\prime}),
\end{equation}
where $\hat{F}(\omega^{\prime}, \mathbf{k}^{\prime})$ is the causal anomalous Green's function. (The normal and anomalous Green functions are defined in Sec.~II in the SM.) As usual, the gap equation follows from the extremum of the MF action given in Eq.~(\ref{model-OF-S-MF-1}) with respect to $\Delta^{\dag}_{MF}$.

To study the possibility of the OF superconductivity in 3D DSM, we employ a minimal model with a single Dirac point. The case of 2D DSM is considered in Sec.~VI in the SM. The explicit form of the low-energy Hamiltonian for the free electrons reads
\begin{equation}
\hat{H}(\mathbf{r}) = -\mu \mathds{1}_4 -iv_F \gamma^0\left(\bm{\gamma}\cdot \bm{\nabla}\right).
\label{model-free-Hamiltonian-Weyl-rel}
\end{equation}
Here $\mu$ is the electric chemical potential, $v_F$ is the Fermi velocity, $\gamma^0$ and $\bm{\gamma}$ are mutually anticommuting gamma matrices. In addition, it is convenient to introduce $\gamma_5 \equiv i\gamma^0\gamma_x\gamma_y\gamma_z$ matrix, whose eigenvalues correspond to chirality $\chi$ degree of freedom.

\begin{figure*}[!ht]
\begin{center}
\includegraphics[width=0.4\textwidth]{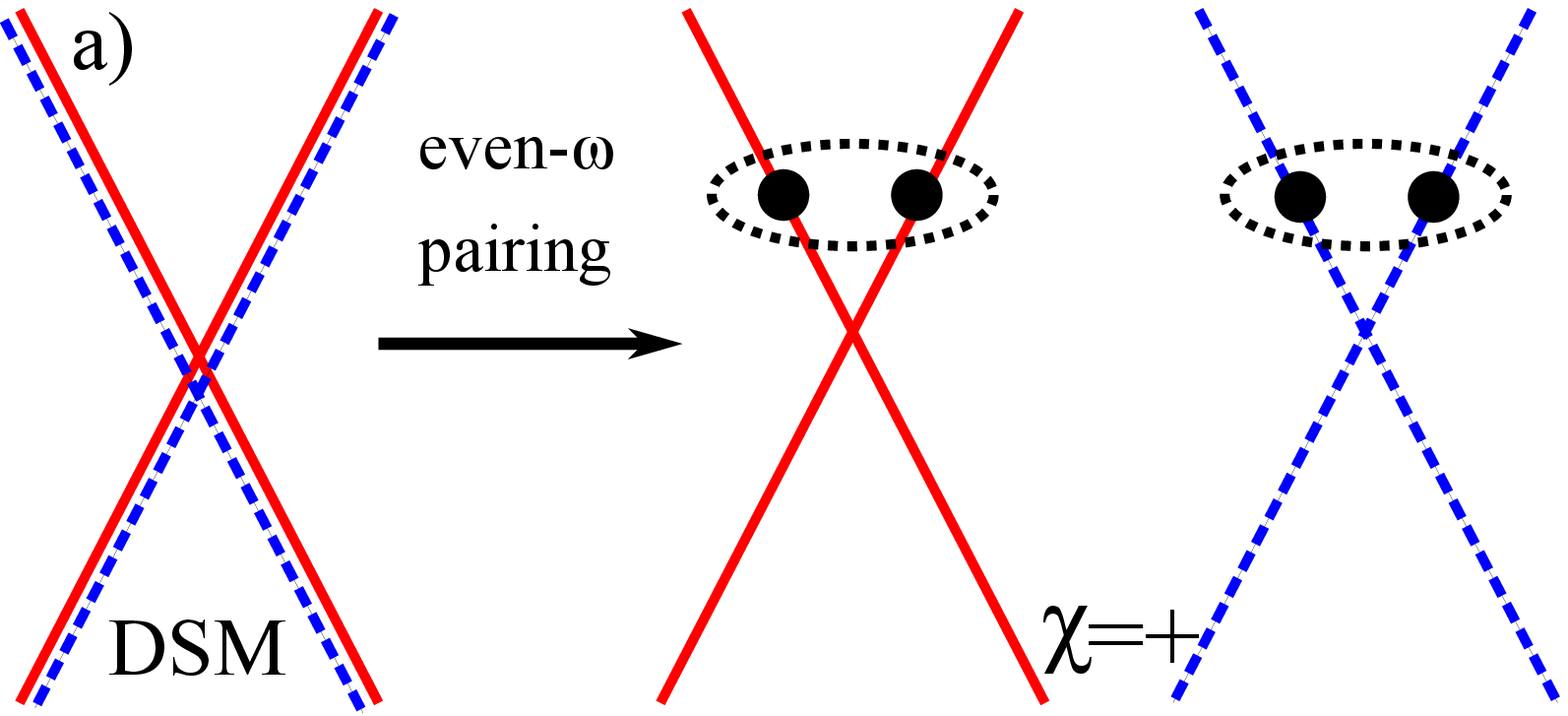}\hfill
\includegraphics[width=0.4\textwidth]{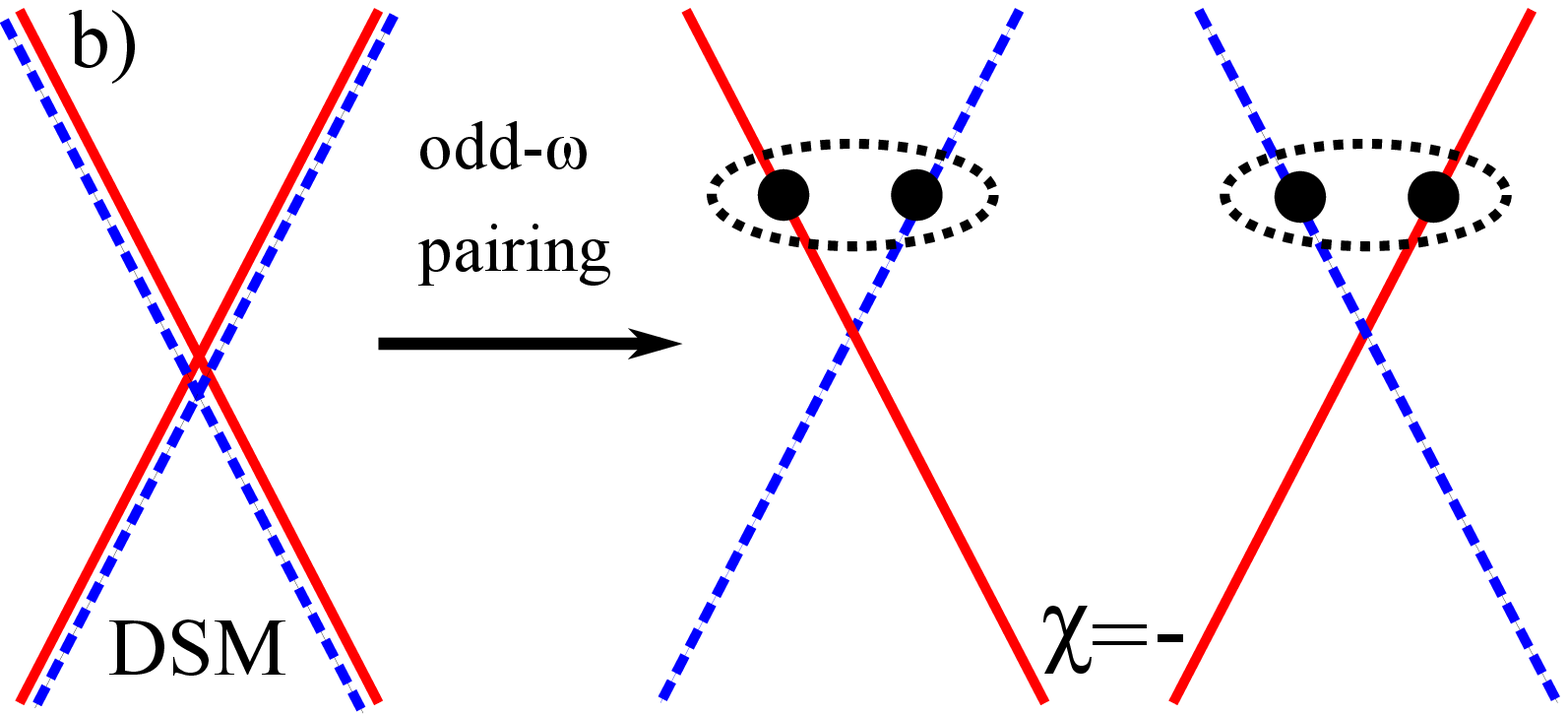}
\end{center}
\caption{The spin-singlet and s-wave pairing channels in Dirac semimetals (DSM) in the case of the even- (panel a)) and odd- (panel b)) frequency pairings. The EF gap corresponds to pairing of the quasiparticles of the same chirality (intra-node channel). On the other hand, the OF Cooper pairs are allowed by the inter-chirality pairing. In both panels, red and blue dashed lines corresponds to the right- and left-handed quasiparticles.
}
\label{fig:Dirac-even-odd}
\end{figure*}

Now, let us discuss the structure of both even- and odd-frequency gaps. In general, two distinctive types of superconducting pairing of chiral fermions can be considered in Weyl and Dirac semimetals~\cite{Meng-Balents:2012,Cho-Moore:2012,Wei:2014vsa,Hosur:2014fba,Bednik:2015tha,Kobayashi2015,Kim-Gilbert:2016,Hashimoto2016}, which are schematically described in Figs.~\ref{fig:Dirac-even-odd}a) and \ref{fig:Dirac-even-odd}b).
For simplicity, we consider the following gaps that are odd and even in frequency:
\begin{eqnarray}
\label{model-OF-gaps-inter-SS-swave}
\hat{\Delta}_{\rm odd}(\omega) = i\sigma_y\otimes\mathds{1}_2 \Delta(\omega),
\end{eqnarray}

\begin{eqnarray}
\label{model-OF-gaps-intra-SS-swave}
\hat{\Delta}_{\rm even}(\omega) = \mathds{1}_2\otimes\mathds{1}_2 \Delta(\omega),
\end{eqnarray}
respectively. The gap properties are summarized in Table~\ref{tab:model-OF-gaps-SPOT}.

\begin{table}[b]
  \centering
\begin{tabular}{c c c c c c}
  \hline
   $\Delta$ & $S$ & $P^{*}$ & $\chi$ & $T^{*}$ & Total\\ \hline
   $\Delta_{\rm odd}(\omega)$ & - & + & - & - & - \\
   $\Delta_{\rm even}(\omega)$ & - & + & + & + & - \\
\end{tabular}
\caption{The odd- and even-frequency gaps given in Eqs.~(\ref{model-OF-gaps-inter-SS-swave}) and (\ref{model-OF-gaps-intra-SS-swave}) as well as their symmetry $SP^{*}\chi T^{*}$ classification. Here $S$, $P^{*}$, $\chi$, $T^{*}$ denote the symmetry properties with respect to the spin, relative coordinate, chirality, and relative time permutations, respectively. The $SP^{*}\chi T^{*}=-1$ rule, which is analogous to the $SP^{*}O T^{*}=-1$ in multiorbital systems~\cite{Linder-Balatsky:rev-2017}, is satisfied for spin-singlet, s-wave, odd-frequency pairing due to the additional chirality degree of freedom. (For details see Sec.~I in the SM.)}
\label{tab:model-OF-gaps-SPOT}
\end{table}

{\em Gap equation.---}
The gap equation (\ref{Gap-Eq-def}) is an integral equation that usually determines an unknown gap function $\Delta(\omega,\mathbf{k})$ for a predefined potential. In the case of the OF superconductivity,  we find it convenient to reformulate the gap equation as an equation for the pairing potential itself. In what follows, a theoretical scheme that is both able to restore the pairing potential via the known gap and determine the gap via the known potential is provided.

For simplicity, we consider only the case of the s-wave pairing, where the dependence on momentum can be omitted, although the approach can be generalized to other cases. Below we concentrate only on the case of vanishing temperature $T\to0$ and compensated state with electric chemical potential $\mu\to0$. In addition, our considerations are restricted only to the case of the pairing potentials that do not grow at $\omega\to\infty$. Then, performing the transformation $\omega\to i\omega$ in Eq.~(\ref{Gap-Eq-def}), one obtains the following gap equation:
\begin{eqnarray}
\label{Gap-Eq-1}
\hat{\Delta}(\omega) = \int_{-\infty}^{\infty} d \omega^{\prime}\hat{V}\left(\omega- \omega^{\prime}\right) \hat{f}(\omega^{\prime}),
\end{eqnarray}
where
\begin{eqnarray}
\label{Gap-Eq-inter-SS-swave-f-def-0}
\hat{f}(\omega^{\prime}) \equiv \int \frac{d^3k}{(2\pi)^4} \hat{F}(\omega^{\prime})
\end{eqnarray}
with the momentum integral taken up to a cutoff $\Lambda_k$. The explicit form of $\hat{f}(\omega^{\prime})$ in the case of odd- and even-frequency pairings is given in Sec.~III in the SM. Since the matrix structure of $\hat{f}(\omega^{\prime})$ coincides with that in $\hat{\Delta}(\omega)$ and $\hat{V} \propto \mathds{1}_8$, the matrix structures will be omitted in what follows.

A convenient method to solve the integral equation (\ref{Gap-Eq-1}) is to transform it into the differential one with the appropriate boundary conditions. As the starting point, the scalar part of the gap equation (\ref{Gap-Eq-1}) can be approximated as follows (see also Sec.~III.A in the SM)
\begin{eqnarray}
\label{Gap-Eq-inter-SS-swave-potential}
\Delta_{\rm odd}(\omega) &=& -2\int_0^{\omega}d\omega^{\prime} \omega^{\prime} V^{\prime}(\omega) f_{\rm odd}(\omega^{\prime}) \nonumber\\
&-& 2\int_{\omega}^{\infty}d\omega^{\prime} \omega V^{\prime}(\omega^{\prime}) f_{\rm odd}(\omega^{\prime}).
\end{eqnarray}
We note that only the derivative with respect to the frequency from the potential $V^{\prime}(\omega)\equiv \partial_{\omega} V(\omega)$ enters the gap equation for the OF gap. Thus, we conclude that the Berezinskii pairing can be supported by a wide range of potentials that have a suitable derivative. The differential form of the gap equation itself is derived in Sec.~III.A in the SM.

The equation for the pairing potential $V(\omega)$ that allows for the OF pairing reads (for the details of the derivation, see Sec.~III.A in the SM)
\begin{eqnarray}
\label{Gap-Eq-inter-SS-swave-potential-diff-All}
&&\omega V^{\prime \prime}(\omega) - V^{\prime}(\omega) = -\frac{\omega\Delta_{\rm odd}^{\prime}(\omega)-\Delta_{\rm odd}(\omega)}{2\int_0^{\omega}d\omega^{\prime} \omega^{\prime} f_{\rm odd}(\omega^{\prime})}.
\end{eqnarray}
The boundary conditions are
\begin{eqnarray}
\label{Gap-Eq-inter-SS-swave-potential-diff-BC}
&&V^{\prime}(\omega)\Big|_{\omega\to\infty} = -\frac{\Delta_{\rm odd}(\omega)}{2\int_0^{\omega}d\omega^{\prime} \omega^{\prime} f_{\rm odd}(\omega^{\prime})} \Big|_{\omega\to\infty},\\
\label{Gap-Eq-potential-crit-V-BC}
&&V(\omega)\Big|_{\omega\to \infty}=0.
\end{eqnarray}
The last condition is crucial for determining the pairing potential and is physically motivated by that fact that the potential should vanish at large frequencies.

Next, let us consider the case of EF pairing. The gap equation (\ref{Gap-Eq-1}) can be approximated as
\begin{eqnarray}
\label{Gap-Eq-even-Delta-1}
\Delta_{\rm even}(\omega) &=& 2\int_0^{\omega}d\omega^{\prime} V(\omega) f_{\rm even}(\omega^{\prime}) \nonumber\\
&+& 2\int_{\omega}^{\infty}d\omega^{\prime} V(\omega^{\prime}) f_{\rm even}(\omega^{\prime}).
\end{eqnarray}
Unlike the case of the OF pairing, the EF gap is sensitive to potential $V(\omega)$ itself. For the details of the derivation as well as the differential gap equation, see Sec.~III.B in the SM.

The equation for the EF pairing potential is
\begin{eqnarray}
\label{Gap-Eq-even-V-eq}
V^{\prime}(\omega) = \frac{\Delta_{\rm even}^{\prime}(\omega)}{2\int_0^{\omega}d\omega^{\prime} f_{\rm even}(\omega^{\prime})},
\end{eqnarray}
with the boundary condition
\begin{eqnarray}
\label{Gap-Eq-even-V-BC}
V(\omega)\Big|_{\omega\to\infty} = \frac{\Delta_{\rm even}(\omega)}{2\int_0^{\omega}d\omega^{\prime} f_{\rm even}(\omega^{\prime})} \Big|_{\omega\to\infty}.
\end{eqnarray}

{\em Pairing potentials.---}
Let us illustrate the proposed framework by calculating the pairing potentials for a few representative gap ansatzes.
The simplest EF gap is
\begin{eqnarray}
\label{Wick-Delta-0}
\Delta_{\rm even}(\omega)= \alpha.
\end{eqnarray}
As for the OF gap, let us consider only the ansatz that produces a vanishing at $\omega\to\infty$ potential,
\begin{eqnarray}
\label{Wick-Delta-2-new}
\Delta_{\rm odd}(\omega)= \alpha \frac{\Lambda_k}{\omega}.
\end{eqnarray}

\begin{figure*}[!ht]
\begin{center}
\includegraphics[height=0.4\textwidth]{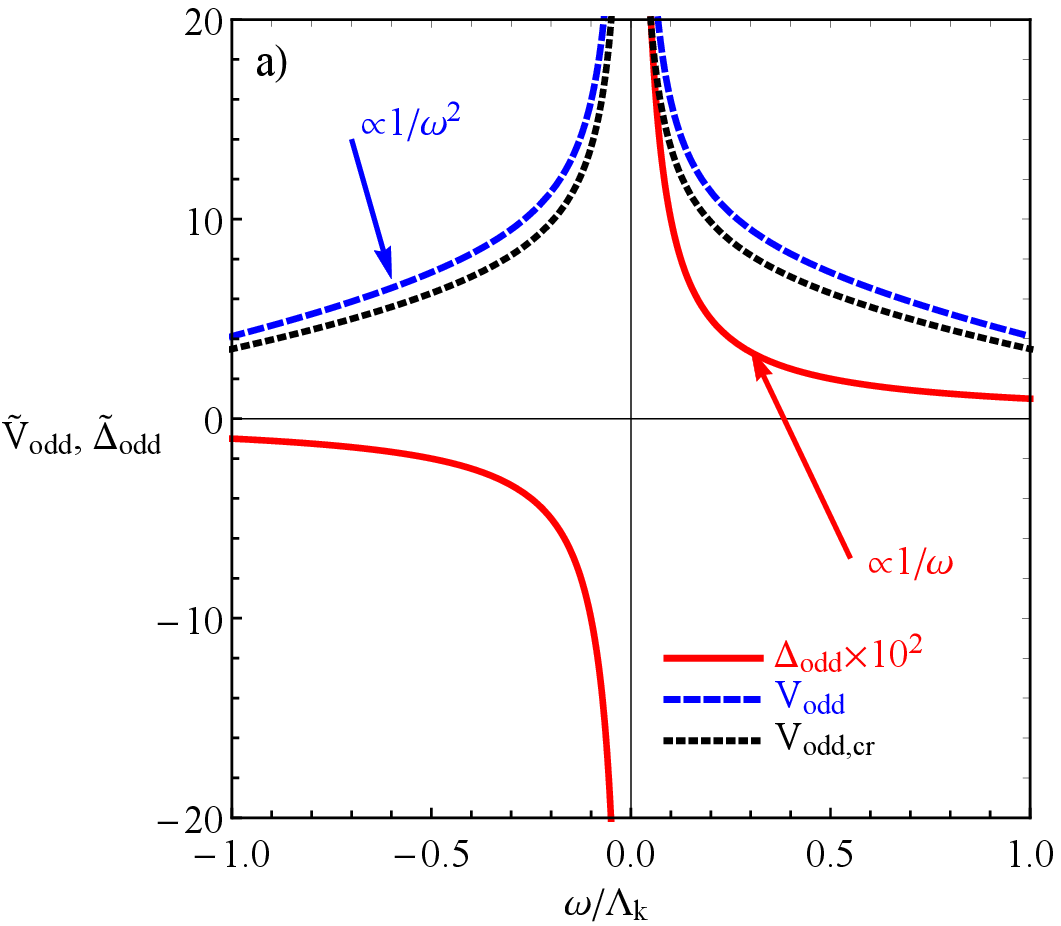}\hfill
\includegraphics[height=0.4\textwidth]{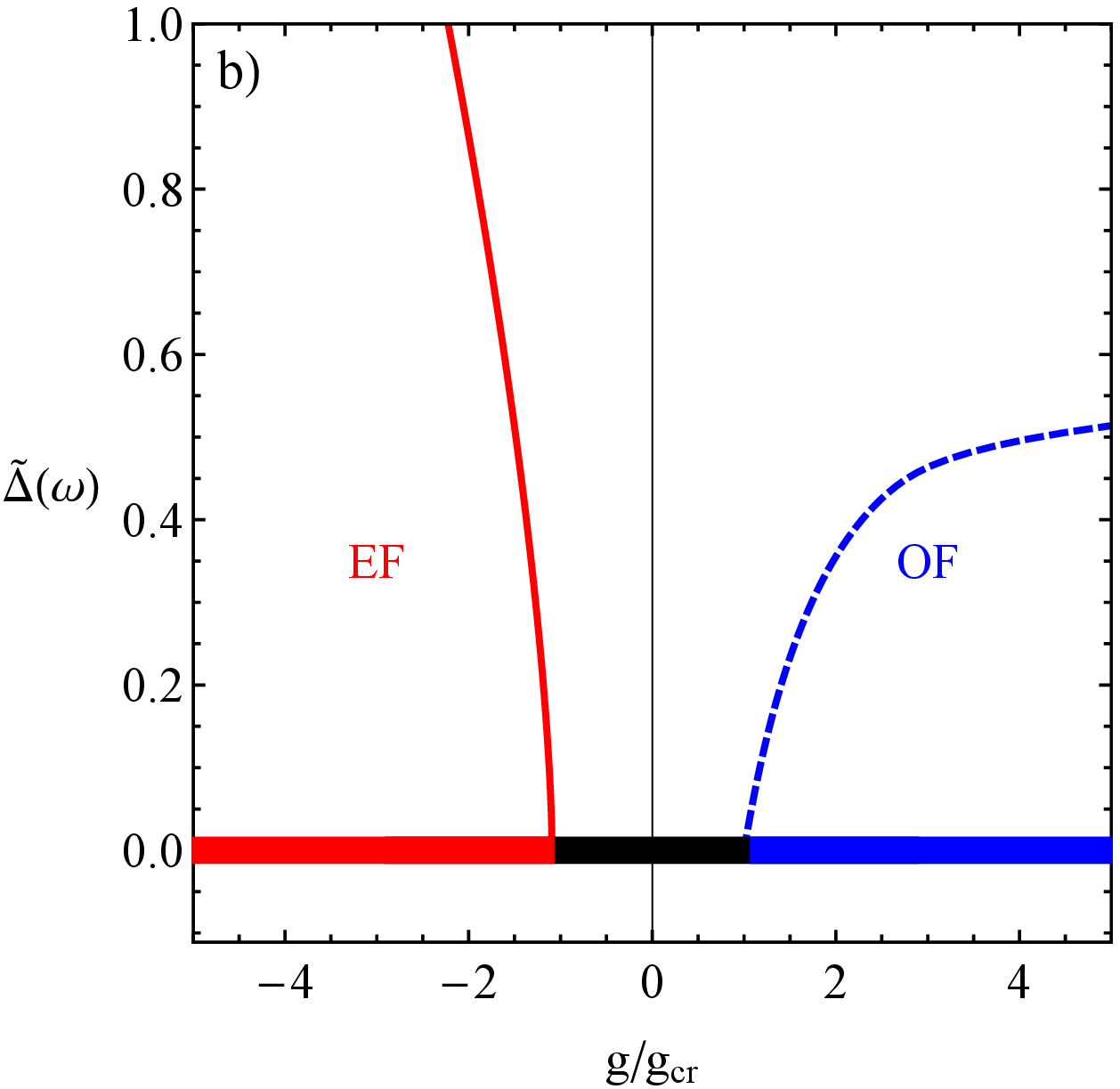}
\end{center}
\caption{Panel a): The odd-frequency gap as well as the corresponding pairing potential as functions of frequency at $\alpha=10^{-2}\Lambda_k$. Red solid lines correspond to the OF gap and blue dashed ones represent the corresponding potential. Black dotted lines represent the critical value of the pairing potential obtained at $\alpha\to0$. Panel b): The dependence of the gaps on the coupling constant for the EF (red solid line) and OF (blue dashed line) gaps.
The 1D phase diagram is shown by the thick red, blue, and black lines at $\tilde{\Delta}=0$.
For the sake of definiteness, $\omega=10^{-2}\Lambda_k$. In addition, $\tilde{\Delta}(\omega)=\Delta(\omega)/\Lambda_k$ and $\tilde{V}(\omega)=V(\omega)/|V_{\rm crit}|$, where $V_{\rm crit}$ is given in Eq.~(\ref{Wick-Delta-0-crit}). Note also that we used a finite frequency cutoff $\Lambda_{\omega}=2\Lambda_k$ in the numerical calculations.
}
\label{fig:Gap-Eq-inter-SS-swave-Delta-even-V}
\end{figure*}

Due to the complicated form of the functions $f_{\rm odd}(\omega)$ and $f_{\rm even}(\omega)$, which is provided in Sec.~III of the SM, the solutions for the potentials are obtained numerically. The OF gap as well as the corresponding pairing potential are presented in Fig.~\ref{fig:Gap-Eq-inter-SS-swave-Delta-even-V}a). We found that the potential diminishes approximately as $\propto 1/\omega^{2}$ at large frequencies and diverges at small ones. In agreement with the prior results (see, e.g., Ref.~\cite{Huang-Balatsky:2015}), we find a critical value of the potential strength  that is needed to generate the OF gap (see the black dotted line in Fig.~\ref{fig:Gap-Eq-inter-SS-swave-Delta-even-V}a)).

Next, let us discuss the dependence of the even- and odd-frequency gaps on the coupling constant $g$. The corresponding results are shown in Fig.~\ref{fig:Gap-Eq-inter-SS-swave-Delta-even-V}b). We found that the potential for the EF gap does not depend on frequency and its critical value, which separates the normal and EF superconducting phases, reads as
\begin{eqnarray}
\label{Wick-Delta-0-crit}
V_{\rm crit} =g_{\rm cr}^{\rm even}= -\frac{8\pi^2 v_F^3}{\Lambda_k^2}.
\end{eqnarray}
As expected, the pairing potential is attractive in this case. Further, the coupling constant should exceed the critical value, i.e., $|g|\geq |g_{\rm cr}^{\rm even}|$, to allow for an EF gap, which is due to the vanishing DOS at the Fermi level.

As for the pairing potential for the OF case, it also should be sufficiently strong to generate the gap, i.e., $g\geq g_{\rm cr}^{\rm odd}$. This critical value is determined at $\alpha\to0$ assuming that the potential $V$ can be factorized as $V(\alpha, \omega)=g(\alpha)\tilde{V}(\omega)$. Then, $g/g_{\rm cr}^{\rm odd} = V(\alpha, \omega)/V(0, \omega)$.
It is worth noting that, in general, such a separation is not exact, which, however, does not change our main qualitative conclusions. The gap as well as the dependence of the corresponding potentials on frequency are summarized in Table~\ref{tab:Delta-V}. Another key result of this study is that the pairing potential for the OF gap (\ref{Wick-Delta-2-new}) can be \emph{repulsive}, $V(\omega)>0$. The results for 2D DSM, which are presented in Sec.~VI in the SM, are qualitatively the same.

\begin{table}[!ht]
  \centering
\begin{tabular}{c c c}
  \hline
   $\Delta(\omega)$ & $V(\omega)$ \\ \hline
   $\alpha$ &  $const<0$ & \\
   $\alpha \Lambda_k/\omega$ & $\propto1/\omega^2 >0$ & \\
\end{tabular}
\caption{The even- and odd-frequency gaps as well as the qualitative dependence of the corresponding potentials on frequency.}
\label{tab:Delta-V}
\end{table}

\begin{figure*}[ht]
\begin{center}
\includegraphics[height=0.4\textwidth]{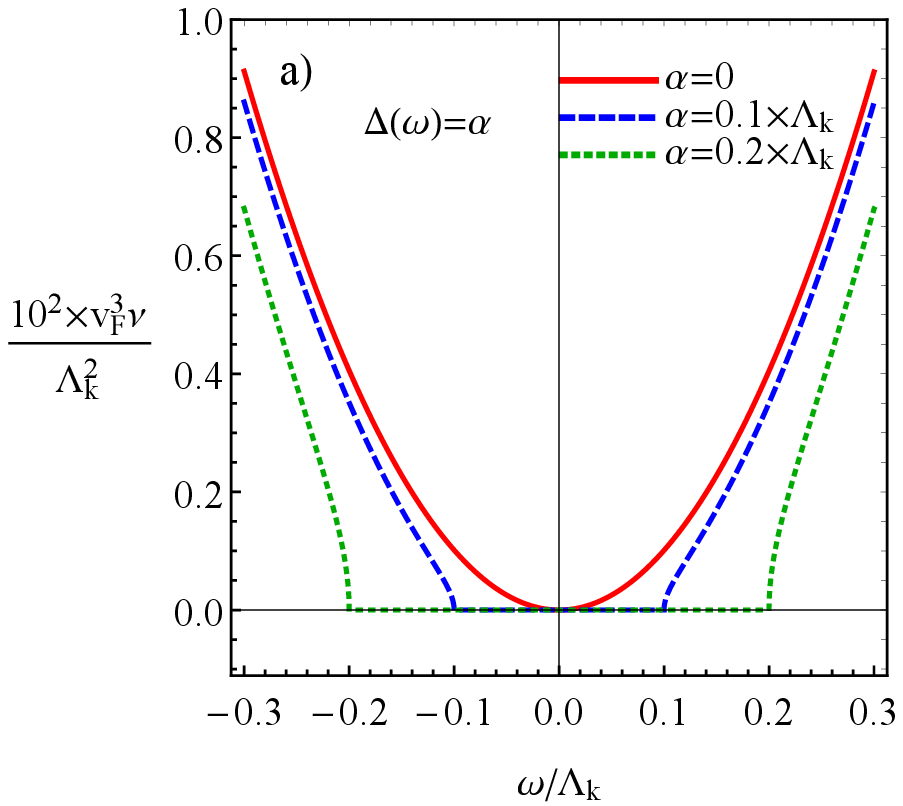}\hfill
\includegraphics[height=0.4\textwidth]{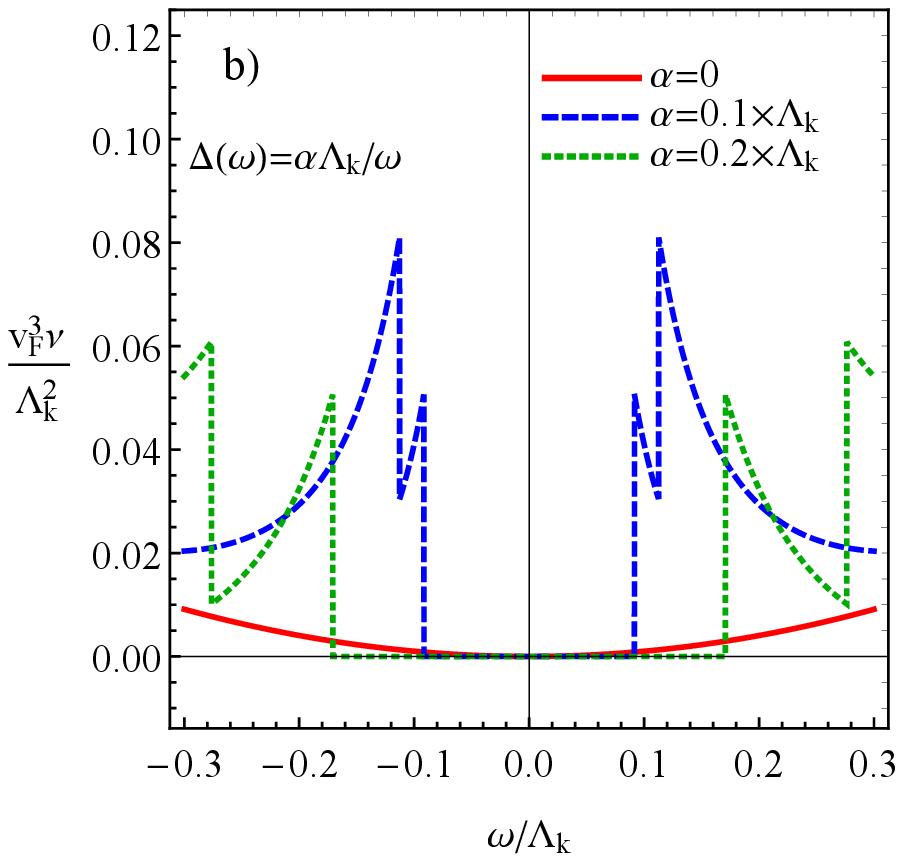}
\end{center}
\caption{The dependence of the electron DOS $\nu(\omega)$ on frequency at a few values of the gap strength $\alpha$ for $\Delta(\omega)=\alpha$ (panel a)) and $\Delta(\omega)=\alpha \Lambda_k/\omega$ (panel b)). The electric chemical potential is $\mu=0$. As expected, the generation of the EF gap pushes the states away from the region of small frequencies, where the spectral weight is recovered at the energy cutoff $\omega\approx \Lambda_k$ (see also Fig.~\ref{fig:App-Observables-DOS-even} in the SM).
The case of the Berezinskii pairing is qualitatively different and is manifested in the formation of the cusp-like features at small $\omega$.
These features originate from the additional branches of the energy dispersion induced by the OF gap. (The details are provided in Sec.~V in the SM.)
}
\label{fig:Observables-DOS}
\end{figure*}

{\em Conclusions.---}
We solved the gap equation for both odd- and even-frequency superconductivity pairings in Dirac semimetals.
By using the effective action approach, we derive the integral gap equation and show how to convert it into a differential one with the appropriate boundary conditions. There are two ways how the proposed framework can be utilized.
The first is to determine the superconductivity gap via a given potential.
The second way is to consider the inverse problem in which the pairing potential is determined via the predefined gap.
We start by noting that both even- and odd-frequency pairings require values of the coupling potential that exceed some critical values in Dirac semimetals at the charge neutrality point.
This allows to rule out the ubiquity of the former and puts both pairings on an equal footing.

We show that the gap equation for an EF pairing is determined by the pairing potential itself. On the other hand, the OF gap depends only on the derivatives from the potential with respect to frequency. Thus, the OF pairing can be generated even by a \emph{repulsive} potential with an appropriate derivative.

The proposed scheme is illustrated for spin-singlet, s-wave gaps.
In agreement with the general consideration, the pairing potential for the OF gap is indeed repulsive. The corresponding derivative, which enters the gap equation, is however, negative.
This finding should be contrasted to the case of the EF pairing, where the gap exists only for an attractive interaction.
Thus, we suggest a new scenario in the search of the superconductors that support Berezinskii pairing due to strongly frequency dependent repulsive pairing potential.

We compare the key aspects of even- and odd-frequency superconductivity in conventional metals and DSM in Table~\ref{tab:even-odd}. As an experimental signature of the Berezinskii pairing, we propose to study the DOS, where characteristic double-cusp features universally appear. The corresponding results are shown and discussed in Figs.~\ref{fig:Observables-DOS}a) and \ref{fig:Observables-DOS}b) (see also the SM).

\begin{table*}[ht]
  \centering
\begin{tabular}{c c c c}
  \hline
    & BCS, even-frequency & Odd-frequency & \\ \hline
   Metal: $\epsilon_k = k^2/(2m)-\mu$ & no critical coupling $g_{\rm crit}\to0$ & critical coupling $|g|>|g_{\rm crit}|$ & \\
   Dirac semimetal: $\epsilon_k = v_F k$  & critical coupling $g<g_{\rm crit}^{\rm even}<0$ & critical coupling $g>g_{\rm crit}^{\rm odd}>0$ & \\
\end{tabular}
\caption{The key aspects of the even- and odd-frequency superconductivity in conventional metals and DSM. Unlike the conventional BCS superconductivity in metals, the generation of both odd- and even-frequency gaps requires the pairing potential to exceed a certain critical value. Further, the OF pairing might be possible for repulsive interactions.}
\label{tab:even-odd}
\end{table*}

While the explicit calculations were performed in the case of the DSM within a spin-singlet and s-wave channel, we believe that the proposed scenario is quite general and can be, in principle, realized in various systems. Such a claim, however, requires further rigorous verification. In addition, the physical nature of repulsive frequency-dependent potential should be also clarified. Further, since the potential is repulsive, there might be some insulating instabilities, which were ignored so far. It would be interesting also to study the renormalization group flows and the critical exponents for the OF transitions in 3D Dirac and Weyl semimetals. Such investigations are, however, beyond the scope of this study and will be reported elsewhere.

\begin{acknowledgments}
We are grateful to E.~Langmann and M.~Geilhufe for useful discussion.
This work was supported by the VILLUM FONDEN via the Centre of Excellence for Dirac Materials (Grant No. 11744), the European Research Council under the European Unions Seventh Framework Program Synergy HERO, and the Knut and Alice Wallenberg Foundation.
\end{acknowledgments}

\begin{widetext}
\newpage
\begin{center}
\textbf{\large \mytitle~--- Supplemental Material} \\[4pt]
P.~O.~Sukhachov,$^{1}$
Vladimir~Juri\v{c}i\'{c},$^{1}$
and A.~V.~Balatsky$^{1,2}$\\[4pt]
$^{1}${\small\it Nordita, KTH Royal Institute of Technology and Stockholm University, Roslagstullsbacken 23, SE-106 91 Stockholm, Sweden}
$^{2}${\small\it Department of Physics, University of Connecticut, Storrs, CT 06269, USA}
\end{center}

\setcounter{page}{1}

\setcounter{equation}{0}
\setcounter{figure}{0}
\renewcommand{\theequation}{S\arabic{equation}}
\renewcommand{\thefigure}{S\arabic{figure}}

\section{I. Berezinskii symmetry relation}
\label{sec:App-Berezinskii}

By following Berezinskii~[S1] and recent advancements discussed, e.g., in Ref.~[S2], we discuss the symmetry properties of the Cooper pairs. The starting point is the two-fermion correlation function $\Delta_{\alpha \beta, ab}(t, \mathbf{r})=\langle \mathcal{T}c_{\alpha, a}(t,\mathbf{r}) c_{\beta, b}(0,\mathbf{0})\rangle$. Here, $\mathcal{T}$ is the time-ordering operator, $c_{\alpha,b}$ is the fermion annihilation operator, $\alpha$ and $\beta$ correspond to the spin indices, $a$ and $b$ denote the orbital or, in the case of Weyl and Dirac semimetals, chirality indices, $t$ is the relative time, and $\mathbf{r}$ is the relative coordinate.
Next, let us define the key symmetry operations. The spin permutation $S$ is
\begin{equation}
\label{App-Berezinskii-S}
S \Delta_{\alpha \beta, ab}(t, \mathbf{r}) S^{-1} = \Delta_{\beta \alpha, ab}(t, \mathbf{r}).
\end{equation}
The relative coordinate inversion $P^{*}$ acts as
\begin{equation}
\label{App-Berezinskii-P}
P^{*} \Delta_{\alpha \beta, ab}(t, \mathbf{r}) (P^{*})^{-1} = \Delta_{\alpha \beta, ab}(t, -\mathbf{r}).
\end{equation}
Further, the chirality indices can be permuted
\begin{equation}
\label{App-Berezinskii-chi}
\chi \Delta_{\alpha \beta, ab}(t, \mathbf{r}) \chi^{-1} = \Delta_{\alpha \beta, ba}(t, \mathbf{r}).
\end{equation}
Finally, the permutation of relative time reads as
\begin{equation}
\label{App-Berezinskii-T}
T^{*} \Delta_{\alpha \beta, ab}(t, \mathbf{r}) (T^{*})^{-1} = \Delta_{\alpha \beta, ab}(-t, \mathbf{r}).
\end{equation}
Note that $P^{*}$ and $T^{*}$ are not the full space-inversion and time-reversal symmetries. For example, $T^{*}$ does not contain the complex conjugation operator.
The two-fermion correlation function $\Delta_{\alpha \beta, ab}(t, \mathbf{r})$ has the following symmetry property under the combined permutation of spin, relative space coordinates, chirality, and relative time:
\begin{equation}
\label{App-Berezinskii-SPOT}
SP^{*}\chi T^{*} \Delta_{\alpha \beta, ab}(t, \mathbf{r}) = -\Delta_{\alpha \beta, ab}(t, \mathbf{r}),
\end{equation}
which is symbolically denoted as $SP^{*}\chi T^{*}=-1$ rule and is analogous to the $SP^{*}O T^{*}=-1$ rule for multiorbital systems~[S2]. This rule plays an important role in classifying possible pairing channels in various superconductors. In the case of Dirac semimetals, the symmetry properties of simplest even- and odd-frequency Cooper pairs are summarized in Table~I in the main text.

\section{II. Effective action approach}
\label{sec:App-model-OF-EA}

By following Refs.~[S3, S4],
let us formulate the mean-field (MF) action for the odd-frequency (OF) or Berezinskii pairing. Note that the corresponding results are also valid for the even-frequency (EF) superconductors.

The partition function of the system is given in terms of the functional integral (see, e.g., Ref.~[S5]):
\begin{eqnarray}
\label{App-model-OF-Z}
Z=\int D\bar{\Psi}D\Psi D\Delta^{\dag} D\Delta e^{-iS(\bar{\Psi},\Psi, \Delta^{\dag}, \Delta)},
\end{eqnarray}
where the action is
\begin{eqnarray}
\label{App-model-OF-S}
S &=& \sum_{\alpha, \beta}\int dx \bar{\Psi}_{\alpha}(x)\left(-i\partial_{t}\delta_{\alpha \beta}+\hat{H}_{\alpha \beta}\right)\Psi_{\beta}(x) + \frac{1}{2} \sum_{\alpha, \beta, \gamma, \delta} \int dx_1 dx_2 \left[V^{-1}(x_1,x_2)\right]_{\alpha \beta \gamma \delta} \Delta^{\dag}_{\alpha \beta}(x_1,x_2) \Delta_{\gamma \delta}(x_1,x_2)\nonumber\\
&+& \frac{1}{2} \sum_{\alpha, \beta} \int dx_1 dx_2 \left[\Delta^{\dag}_{\alpha \beta}(x_1,x_2)\Psi_{\alpha}(x_2)\Psi_{\beta}(x_1) +\Delta_{\alpha \beta}(x_1,x_2)\bar{\Psi}_{\alpha}(x_1)\bar{\Psi}_{\beta}(x_2) \right].
\end{eqnarray}
Here $x=\left(t,\mathbf{r}\right)$ is the time-space four-vector, $\Psi_{\alpha}(x)$ is the fermionic field, $\bar{\Psi}_{\alpha}(x)$ is the Dirac conjugate of the fermionic field, $\Delta_{\alpha \beta}(x_1,x_2)$ is the pairing field, and $V(x_1-x_2)$ is the interaction potential. Further, indices $\alpha$, $\beta$, $\gamma$, $\delta$ denote the components of the fermion fields. The pairing field was introduced by Hubbard--Stratonovich transformation via the decoupling of the interaction term (see, e.g., Ref.~[S6])
\begin{equation}
\label{App-model-OF-V-term}
\sum_{\alpha, \beta, \gamma, \delta} \int dx_1 dx_2 V_{\alpha \beta \gamma \delta}(x_1-x_2) \bar{\Psi}_{\alpha}(x_1)\bar{\Psi}_{\beta}(x_2) \Psi_{\gamma}(x_2)\Psi_{\delta}(x_1).
\end{equation}
In what follows, we will omit explicit indices at fermion and pairing fields as well as the interaction potential for the sake of the notations clarity.

In the mean-field approximation, the normal and anomalous Green's functions are defined as
\begin{eqnarray}
\label{App-model-OF-G}
G_{\alpha \beta}(t-t^{\prime}; \mathbf{r}-\mathbf{r}^{\prime}) = -i Z_{MF}^{-1} \int D\bar{\Psi}D\Psi\, \Psi_{\alpha}(t, \mathbf{r})\bar{\Psi}_{\beta}(t^{\prime}, \mathbf{r}^{\prime}) e^{-iS_{MF}},\\
\label{App-model-OF-tG}
\tilde{G}_{\alpha \beta}(t-t^{\prime}; \mathbf{r}-\mathbf{r}^{\prime}) = -i Z_{MF}^{-1} \int D\bar{\Psi}D\Psi\, \bar{\Psi}_{\alpha}(t, \mathbf{r})\Psi_{\beta}(t^{\prime}, \mathbf{r}^{\prime}) e^{-iS_{MF}},
\end{eqnarray}
and
\begin{eqnarray}
\label{App-model-OF-F}
F_{\alpha \beta}(t-t^{\prime}; \mathbf{r}-\mathbf{r}^{\prime}) = i Z_{MF}^{-1} \int D\bar{\Psi}D\Psi\, \Psi_{\alpha}(t, \mathbf{r})\Psi_{\beta}(t^{\prime}, \mathbf{r}^{\prime}) e^{-iS_{MF}},\\
\label{App-model-OF-F-dag}
F_{\alpha \beta}^{\dag}(t-t^{\prime}; \mathbf{r}-\mathbf{r}^{\prime}) = i Z_{MF}^{-1} \int D\bar{\Psi}D\Psi\, \bar{\Psi}_{\beta}(t, \mathbf{r})\bar{\Psi}_{\alpha}(t^{\prime}, \mathbf{r}^{\prime}) e^{-iS_{MF}},
\end{eqnarray}
respectively. The corresponding mean-field action $S_{MF}$ is
\begin{eqnarray}
\label{App-model-OF-S-MF}
S_{MF} &=& \int dx \bar{\Psi}(x)\left[-i\partial_{t}+\hat{H}(x)\right]\Psi(x) +\frac{1}{2}\int dx_1 dx_2 \left[\Delta^{\dag}_{MF}(x_1-x_2)\Psi(x_2)\Psi(x_1) +\Delta_{MF}(x_1-x_2)\bar{\Psi}(x_1)\bar{\Psi}(x_2) \right] \nonumber\\
&+&\frac{1}{2} \int dx_1 dx_2 \frac{|\Delta_{MF}(x_1-x_2)|^2}{V(x_1-x_2)}.
\end{eqnarray}
Note that the MF order parameter $\Delta_{MF}(x_1-x_2)$ now depends only on the relative coordinate. Further, the MF action (\ref{App-model-OF-S-MF}) can be rewritten as
\begin{eqnarray}
\label{App-model-OF-S-MF-1}
S_{MF} = -\frac{1}{2}\int dx_1 dx_2 \bar{\Psi}_{\rm N}(x_1) G^{-1}_{\rm N}(x_1-x_2) \Psi_{\rm N}(x_2) +\frac{1}{2} \int dx_1 dx_2 \frac{|\Delta_{MF}(x_1-x_2)|^2}{V(x_1-x_2)},
\end{eqnarray}
where
\begin{eqnarray}
\label{App-model-OF-G-1-def}
G^{-1}_{\rm N}(x_1-x_2) = \left[i\partial_{t_1}  -\hat{H}_{\rm N}(x_1)\right] \delta(x_1-x_2) - \left(
                                                              \begin{array}{cc}
                                                                0 & \Delta_{MF}(x_1-x_2)  \\
                                                                \Delta^{\dag}_{MF}(x_1-x_2) & 0 \\
                                                              \end{array}
                                                            \right)
\end{eqnarray}
is the Green function in the Nambu space and Hamiltonian $\hat{H}_{\rm N}$ is given by
\begin{equation}
\label{App-model-OF-HBdG}
\hat{H}_{\rm N}(\mathbf{r}) = \left(
                    \begin{array}{cc}
                      \hat{H}(\mathbf{r}) & 0 \\
                      0 & -\hat{\Theta}\hat{H}(\mathbf{r})\hat{\Theta}^{-1} \\
                    \end{array}
                  \right).
\end{equation}
Here $\hat{\Theta}= i \mathds{1}_2\otimes\sigma_y \hat{K}$ is the time-reversal (TR) operator, $\mathds{1}_2$ is the unit $2\times2$ matrix, $i\sigma_y $ is the Pauli matrix that describes the spin flip, and $\hat{K}$ is the complex conjugation operator.

In the case of 3D Dirac semimetals (DSM), the explicit form of the minimal low-energy Hamiltonian for the free electrons near the $\Gamma$ point in the Brillouin zone reads
\begin{equation}
\hat{H}(\mathbf{r}) = -\mu \mathds{1}_4 -iv_F \gamma^0\left(\bm{\gamma}\cdot \bm{\nabla}\right),
\label{App-model-free-Hamiltonian-Weyl-rel}
\end{equation}
where $\mu$ is the electric chemical potential, $v_F$ is the Fermi velocity, $\gamma^0$ and $\bm{\gamma}$ are gamma matrices.
The explicit form of the latter in the chiral representation reads
\begin{equation}
\gamma^0 = \left( \begin{array}{cc} 0 & -\mathds{1}_2\\ -\mathds{1}_2 & 0 \end{array} \right),\qquad
\bm{\gamma} = \left( \begin{array}{cc} 0& \bm{\sigma} \\  - \bm{\sigma} & 0 \end{array} \right).
\label{App-model-Dirac-matrices}
\end{equation}
In addition, it is convenient to introduce $\gamma_5$ matrix, i.e.,
\begin{equation}
\label{App-model-gamma-5}
\gamma_5 \equiv i\gamma^0\gamma_x\gamma_y\gamma_z = \left( \begin{array}{cc} \mathds{1}_2 & 0\\ 0 & -\mathds{1}_2 \end{array} \right),
\end{equation}
whose eigenvalues correspond to chirality $\chi$ degree of freedom.

By performing the Fourier transform in Eq.~(\ref{App-model-OF-G-1-def}), we obtain
\begin{eqnarray}
\label{App-model-OF-G-1-momentum}
G_{\rm N}(\omega, \mathbf{k}) &=&
 \left[\omega \mathds{1}_8  -\hat{H}_{\rm N}(\mathbf{k})\right] - \left(
                                                              \begin{array}{cc}
                                                                0 & \hat{\Delta}_{MF}(\omega, \mathbf{k})  \\
                                                                \hat{\Delta}^{\dag}_{MF}(\omega, \mathbf{k}) & 0 \\
                                                              \end{array}
                                                            \right) 
                                                            =
                                                            \left(
                                                                                        \begin{array}{cc}
                                                                                          \hat{G}(\omega,\mathbf{k}) & \hat{F}(\omega,\mathbf{k}) \\
                                                                                          \left[\hat{F}(\omega,\mathbf{k})\right]^{\dag} & \hat{\tilde{G}}(\omega,\mathbf{k}) \\
                                                                                        \end{array}
                                                                                      \right).
\end{eqnarray}

In the case of the 3D DSM, whose Hamiltonian is given in Eq.~(\ref{App-model-free-Hamiltonian-Weyl-rel}), the $8\times 8$ Green's function in Eq.~(\ref{App-model-OF-G-1-momentum}) acts in the space of the Nambu--Gor'kov spinors
\begin{equation}
\label{App-OP-Weyl-model-Nambu-Psi}
\Psi_{\rm N}(\mathbf{k})=\left\{\Psi(\mathbf{k}),\Psi_{\Theta}(\mathbf{k})\right\}^{T},
\end{equation}
where
\begin{equation}
\label{App-OP-Weyl-model-Nambu-Psi-1}
\Psi(\mathbf{k}) = \left\{\psi^{\chi=+}_{\uparrow} (\mathbf{k}),\psi^{\chi=+}_{\downarrow}(\mathbf{k}),\psi^{\chi=-}_{\uparrow}(\mathbf{k}),\psi^{\chi=-}_{\downarrow}(\mathbf{k})\right\}^{T}
\end{equation}
and the TR conjugate spinor is given by
\begin{equation}
\label{App-OP-Weyl-model-Nambu-Psi-2}
\Psi_{\Theta}(\mathbf{k}) = \left\{\psi^{\chi=+}_{\downarrow}(-\mathbf{k}),-\psi^{\chi=+}_{\uparrow}(-\mathbf{k}),\psi^{\chi=-}_{\downarrow}(-\mathbf{k}), -\psi^{\chi=-}_{\uparrow}(-\mathbf{k})\right\}^{\dag}.
\end{equation}
Here $\uparrow$ and $\downarrow$ correspond to the states with (pseudo-)spin up and down, respectively.

Further, let us present Green's function in the Nambu space for even- and odd-frequency gaps used in the main text. In the case of the intra-node, spin-singlet, s-wave, and even-frequency gap
\begin{eqnarray}
\label{App-model-OF-gaps-intra-SS-swave}
\hat{\Delta}_{\rm even}(\omega) = \mathds{1}_2\otimes\mathds{1}_2 \Delta(\omega)
= \left(
                                                                       \begin{array}{cc}
                                                                         \Delta(\omega) \mathds{1}_2 & 0 \\
                                                                         0 & \Delta(\omega) \mathds{1}_2 \\
                                                                       \end{array}
                                                                     \right),
\end{eqnarray}
the normal and anomalous Green's functions read
\begin{eqnarray}
\label{App-OP-Weyl-Green-G-intra-SS-swave-even}
\hat{G}_{\rm even}(\omega, \mathbf{k}) &=& \frac{\left[(\omega+\mu)^2-|\Delta(\omega)|^2-v_F^2k^2\right]\left[(\omega-\mu)\mathds{1}_{4} + v_F\gamma^0 (\bm{\gamma}\cdot\mathbf{k})\right] -2\mu|\Delta(\omega)|^2\mathds{1}_{4}}{\tilde{\omega}_{+}^2\tilde{\omega}_{-}^2},\\
\label{App-OP-Weyl-Green-tG-intra-SS-swave-even}
\hat{\tilde{G}}_{\rm even}(\omega, \mathbf{k}) &=& \frac{\left[(\omega+\mu)^2-|\Delta(\omega)|^2-v_F^2k^2\right]\left[(\omega+\mu)\mathds{1}_{4} -v_F\gamma^0 (\bm{\gamma}\cdot\mathbf{k})\right] +2\mu|\Delta(\omega)|^2\mathds{1}_{4} }{\tilde{\omega}_{+}^2\tilde{\omega}_{-}^2},\\
\label{App-OP-Weyl-Green-F-intra-SS-swave-even}
\hat{F}_{\rm even}(\omega, \mathbf{k}) &=& \Delta(\omega)\frac{\left(\omega^2-\mu^2 -|\Delta(\omega)|^2-v_F^2k^2\right) \mathds{1}_4 +2v_F\mu\gamma^0(\bm{\gamma}\cdot\mathbf{k})}{\tilde{\omega}_{+}^2\tilde{\omega}_{-}^2},
\end{eqnarray}
where $\tilde{\omega}_{\pm}=\sqrt{\omega^2-|\Delta(\omega)|^2-(\mu\pm v_Fk)^2}$.

For the spin-singlet, s-wave, inter-node, and odd-frequency gap
\begin{eqnarray}
\label{App-model-OF-gaps-inter-SS-swave}
\hat{\Delta}_{\rm odd}(\omega) = i\sigma_y\otimes\mathds{1}_2 \Delta(\omega)
= \left(
                                                                       \begin{array}{cc}
                                                                         0 & \Delta(\omega) \mathds{1}_2 \\
                                                                         -\Delta(\omega)\mathds{1}_2 & 0 \\
                                                                       \end{array}
                                                                     \right)
\end{eqnarray}
the normal and anomalous Green's functions are
\begin{eqnarray}
\label{App-OP-Weyl-Green-G-inter-SS-swave}
\hat{G}_{\rm odd}(\omega, \mathbf{k}) &=& \frac{(\omega-\mu)\left[(\omega+\mu)^2-v_F^2k^2-|\Delta(\omega)|^2\right]\mathds{1}_4 -2\mu|\Delta(\omega)|^2\mathds{1}_4  +\left[(\omega+\mu)^2-v_F^2k^2+|\Delta(\omega)|^2\right] v_F\gamma^0(\bm{\gamma}\cdot\mathbf{k})}{\left(v_F^2k^2 -\omega_{+}^2\right)\left(v_F^2k^2 -\omega_{-}^2\right)},\nonumber\\
\\
\label{App-OP-Weyl-Green-tG-inter-SS-swave}
\hat{\tilde{G}}_{\rm odd}(\omega, \mathbf{k}) &=& \frac{(\omega+\mu)\left[(\omega-\mu)^2-v_F^2k^2-|\Delta(\omega)|^2\right]\mathds{1}_4 +2\mu|\Delta(\omega)|^2 \mathds{1}_4 -\left[(\omega-\mu)^2-v_F^2k^2+|\Delta(\omega)|^2\right] v_F\gamma^0(\bm{\gamma}\cdot\mathbf{k})}{\left(v_F^2k^2 -\omega_{+}^2\right)\left(v_F^2k^2 -\omega_{-}^2\right)},\nonumber\\
\\
\label{App-OP-Weyl-Green-F-inter-SS-swave}
\hat{F}_{\rm odd}(\omega, \mathbf{k}) &=& \gamma^0\gamma_5 \Delta(\omega) \frac{\left(\omega_{+}\omega_{-}+v_F^2k^2\right)\mathds{1}_4 -2v_F\omega \gamma^0(\bm{\gamma}\cdot\mathbf{k})}{\left(v_F^2k^2 -\omega_{+}^2\right)\left(v_F^2k^2 -\omega_{-}^2\right)},
\end{eqnarray}
where we introduced the shorthand notation $\omega_{\pm}=\omega\pm\sqrt{|\Delta(\omega)|^2+\mu^2}$.

\section{III. Derivation of the odd-frequency gap equation}
\label{sec:App-GE-derive}

Let us present the details of the derivation of the gap equation. To start with, let us assume that the pairing potentials for both even- and odd-frequency pairings do not grow at $\omega\to\infty$. In addition, the dependence of the momentum can be also neglected for the s-wave pairing. Then, by performing the Wick rotation $\omega\to i\omega$, one can obtain the following gap equation:
\begin{eqnarray}
\label{App-Wick-GE}
\hat{\Delta}(\omega) = \int_{-\infty}^{\infty} d \omega^{\prime}\hat{V}\left(\omega- \omega^{\prime}\right) \hat{f}(\omega^{\prime}).
\end{eqnarray}
In what follows, we consider the cases of the odd- and even-frequency pairings separately.

\subsection{III.A Odd-frequency pairing}
\label{sec:App-GE-derive-odd}

Let us start from the case of OF pairing. Since the pairing potentials are even in $\omega-\omega^{\prime}$, the gap equation (\ref{App-Wick-GE}) takes the following form:
\begin{eqnarray}
\label{App-Gap-Eq-potential-GE}
\hat{\Delta}_{\rm odd}(\omega) =  \int_{0}^{\infty} d \omega^{\prime}\left[\hat{V}\left(\omega- \omega^{\prime}\right)-\hat{V}\left(\omega+ \omega^{\prime}\right)\right] \hat{f}_{\rm odd}(\omega^{\prime}).
\end{eqnarray}
Here function $\hat{f}_{\rm odd}(\omega^{\prime})$ on the right-hand side is odd in $\omega^{\prime}$ and reads as
\begin{eqnarray}
\label{App-Wick-f-odd-def}
\hat{f}_{\rm odd}(\omega^{\prime}) &=& \int \frac{d^3k}{(2\pi)^4} \hat{F}_{\rm odd}(\omega^{\prime}) =  \gamma^0\gamma_5\frac{2\Delta_{\rm odd}(\omega^{\prime})}{(2\pi)^3} \int_{0}^{\Lambda_k/v_F} k^2dk \frac{v_F^2k^2- (\omega^{\prime})^2-|\Delta_{\rm odd}(\omega^{\prime})|^2}{\left[(\omega^{\prime})^2+\left(v_Fk+|\Delta_{\rm odd}(\omega^{\prime})|\right)^2\right]\left[(\omega^{\prime})^2+\left(v_Fk-|\Delta_{\rm odd}(\omega^{\prime})|\right)^2\right]} \nonumber\\
&=& \gamma^0\gamma_5\frac{\Delta_{\rm odd}(\omega^{\prime})}{4\pi^3 v_F^3} \Bigg\{\Lambda_k +\omega^{\prime} \left[\arctan{\left(\frac{|\Delta_{\rm odd}(\omega^{\prime})|-\Lambda_k}{\omega^{\prime}}\right)} -\arctan{\left(\frac{|\Delta_{\rm odd}(\omega^{\prime})|+\Lambda_k}{\omega^{\prime}}\right)}\right] \nonumber\\
&+&\frac{|\Delta_{\rm odd}(\omega^{\prime})|^2-(\omega^{\prime})^2}{4|\Delta_{\rm odd}(\omega^{\prime})|} \ln{\left[\frac{(\omega^{\prime})^2+(|\Delta_{\rm odd}(\omega^{\prime})|-\Lambda_k)^2}{(\omega^{\prime})^2+(|\Delta_{\rm odd}(\omega^{\prime})|+\Lambda_k)^2}\right]}
\Bigg\},
\end{eqnarray}
where we set $\mu\to0$ for simplicity. Since the matrix structure is the same on both sides of Eq.~(\ref{App-Gap-Eq-potential-GE}), henceforth, we omit it.

To proceed with the analytical analysis, let us rewrite the potential in the following form:
\begin{eqnarray}
\label{App-Gap-Eq-potential-V-V}
&&V\left(\omega- \omega^{\prime}\right)-V\left(\omega+ \omega^{\prime}\right) \approx -2\left[\theta(\omega-\omega^{\prime}) \omega^{\prime} V^{\prime}(\omega) +\theta(\omega^{\prime}-\omega) \omega V^{\prime}(\omega^{\prime})\right].
\end{eqnarray}
Here $ V^{\prime}(\omega)\equiv\partial_{\omega}V(\omega)$ is the derivative from the potential with respect to $\omega$.
This approximation allows one to simplify the gap equation (\ref{App-Gap-Eq-potential-GE}), i.e.,
\begin{eqnarray}
\label{App-Gap-Eq-inter-SS-swave-potential}
\Delta_{\rm odd}(\omega) = -2\left[\int_0^{\omega}d\omega^{\prime} \omega^{\prime} V^{\prime}(\omega) f_{\rm odd}(\omega^{\prime}) +\int_{\omega}^{\infty}d\omega^{\prime} \omega V^{\prime}(\omega^{\prime}) f_{\rm odd}(\omega^{\prime})\right].
\end{eqnarray}
The next step is to rewrite the integral equation (\ref{App-Gap-Eq-inter-SS-swave-potential}) in terms of the differential equation. Let us first take the derivative with respect to $\omega$,
\begin{eqnarray}
\label{App-Gap-Eq-inter-SS-swave-potential-diff-1}
\Delta^{\prime}_{\rm odd}(\omega) -\frac{\Delta_{\rm odd}(\omega)}{\omega} = -2\int_0^{\omega}d\omega^{\prime} \omega^{\prime} f_{\rm odd}(\omega^{\prime}) \left[V^{\prime \prime}(\omega) -\frac{V^{\prime}(\omega)}{\omega}\right].
\end{eqnarray}
By taking another derivative with respect to $\omega$ in Eq.~(\ref{App-Gap-Eq-inter-SS-swave-potential-diff-1}), one obtains
\begin{eqnarray}
\label{App-Gap-Eq-inter-SS-swave-potential-diff-2}
&&\Delta_{\rm odd}^{\prime\prime}(\omega) -\frac{\Delta_{\rm odd}^{\prime}(\omega)}{\omega} +\frac{\Delta_{\rm odd}(\omega)}{\omega^2} = -2\omega f_{\rm odd}(\omega) \left[V^{\prime \prime}(\omega) -\frac{V^{\prime}(\omega)}{\omega}\right] \nonumber\\
&&-2\int_0^{\omega}d\omega^{\prime} \omega^{\prime} f_{\rm odd}(\omega^{\prime}) \left[V^{\prime \prime \prime}(\omega) -\frac{V^{\prime \prime}(\omega)}{\omega} +\frac{V^{\prime}(\omega)}{\omega^2}\right].
\end{eqnarray}
By using Eq.~(\ref{App-Gap-Eq-inter-SS-swave-potential-diff-1}), we obtain the following gap equation, which is presented in the main text:
\begin{eqnarray}
\label{App-Gap-Eq-inter-SS-swave-potential-diff-fin}
\omega^2 \Delta_{\rm odd}^{\prime\prime}(\omega) -\left[\omega \Delta_{\rm odd}^{\prime}(\omega) -\Delta(\omega)\right]\left[1 + \frac{\omega^2 V^{\prime \prime \prime}(\omega) -\omega V^{\prime \prime}(\omega) +V^{\prime}(\omega)}{\omega V^{\prime \prime}(\omega) -V^{\prime}(\omega)} \right] 
= -2f_{\rm odd}(\omega) \left[\omega V^{\prime \prime}(\omega) -V^{\prime}(\omega)\right].
\end{eqnarray}

Since the corresponding gap equation is the differential equation of the second order, one needs to impose the appropriate boundary conditions. By taking the limit $\omega\to0$ in Eq.~(\ref{App-Gap-Eq-inter-SS-swave-potential-diff-1}), the following relation can be derived:
\begin{eqnarray}
\label{App-Gap-Eq-inter-SS-swave-potential-diff-BC-1}
\left[\Delta_{\rm odd}^{\prime}(\omega) -\frac{\Delta_{\rm odd}(\omega)}{\omega}\right]\left[V^{\prime \prime}(\omega) -\frac{V^{\prime}(\omega)}{\omega}\right]^{-1} \Big|_{\omega\to0} =0.
\end{eqnarray}
The other boundary condition requires us to take the limit $\omega\to\infty$ in Eqs.~(\ref{App-Gap-Eq-inter-SS-swave-potential}) and (\ref{App-Gap-Eq-inter-SS-swave-potential-diff-1}). The resulting expression reads
\begin{eqnarray}
\label{App-Gap-Eq-inter-SS-swave-potential-diff-BC-2}
\Delta_{\rm odd}^{\prime}(\omega)\Big|_{\omega\to\infty}  -\frac{\Delta_{\rm odd}(\omega)}{\omega}\left\{1+\frac{\omega}{V^{\prime}(\omega)} \left[V^{\prime \prime}(\omega) -\frac{V^{\prime}(\omega)}{\omega}\right] \right\} \Big|_{\omega\to\infty} =0.
\end{eqnarray}

Next, let us derive the equation for the pairing potential. Since the potential does not depend on $\omega^{\prime}$ on the right-hand side in Eq.~(\ref{App-Gap-Eq-inter-SS-swave-potential-diff-1}), this equation can be straightforwardly rewritten  as
\begin{eqnarray}
\label{App-Gap-Eq-inter-SS-swave-potential-diff-All}
\omega V^{\prime \prime}(\omega) - V^{\prime}(\omega) = \left[\omega\Delta_{\rm odd}^{\prime}(\omega)-\Delta_{\rm odd}(\omega)\right] \left[-2\int_0^{\omega}d\omega^{\prime} \omega^{\prime} f_{\rm odd}(\omega^{\prime})\right]^{-1}.
\end{eqnarray}
The boundary condition follows from Eq.~(\ref{App-Gap-Eq-inter-SS-swave-potential}), where the limit $\omega\to\infty$ is taken, i.e.,
\begin{eqnarray}
\label{App-Gap-Eq-inter-SS-swave-potential-diff-BC}
V^{\prime}_{\rm odd}(\omega)\Big|_{\omega\to\infty} = \Delta_{\rm odd}(\omega) \left[-2\int_0^{\omega}d\omega^{\prime} \omega^{\prime} f_{\rm odd}(\omega^{\prime})\right]^{-1} \Big|_{\omega\to\infty}.
\end{eqnarray}
It is important to note that since the OF gap equation depends only on the frequency derivatives from the pairing potential, an additional boundary condition should be imposed in order to determine the potential itself. For example, one can require the pairing potential to be vanishing at $\omega\to\infty$
\begin{eqnarray}
\label{App-Gap-Eq-potential-crit-V-BC}
V(\omega)\Big|_{\omega\to \infty}=0,
\end{eqnarray}
which is consistent with our initial assumption that $V(\omega)$
should not grow with $\omega$.

\subsection{III.B Even-frequency pairing}
\label{sec:App-GE-derive-even}

Next, we derive the gap equation in the case of EF pairing. Similarly to Eq.~(\ref{App-Gap-Eq-potential-GE}), the corresponding gap equation reads as
\begin{eqnarray}
\label{App-Gap-Eq-even-GE}
\hat{\Delta}_{\rm even}(\omega) =  \int_{0}^{\infty} d \omega^{\prime}\left[\hat{V}\left(\omega- \omega^{\prime}\right)+\hat{V}\left(\omega+ \omega^{\prime}\right)\right] \hat{f}_{\rm even}(\omega^{\prime}),
\end{eqnarray}
where $\hat{f}_{\rm even}(\omega^{\prime})$ is an even function of $\omega^{\prime}$, whose explicit form in the limit $\mu\to0$ reads as
\begin{eqnarray}
\label{App-Wick-f-even-def}
\hat{f}_{\rm even}(\omega^{\prime}) &=& \int \frac{d^3k}{(2\pi)^4} \hat{F}_{\rm even}(\omega^{\prime}) = -\frac{2\Delta_{\rm even}(\omega^{\prime})}{(2\pi)^3} \int_{0}^{\Lambda_k/v_F} k^2dk \frac{1}{(\omega^{\prime})^2+|\Delta_{\rm even}(\omega^{\prime})|^2+v_F^2k^2} \nonumber\\
&=& - \frac{\Delta_{\rm even}(\omega^{\prime}) \Lambda_k}{4\pi^3 v_F^3} \left[ 1 -\frac{\sqrt{(\omega^{\prime})^2+|\Delta_{\rm even}(\omega^{\prime})|^2}}{\Lambda_k} \arctan{\left(\frac{\Lambda_k}{\sqrt{(\omega^{\prime})^2+|\Delta_{\rm even}(\omega^{\prime})|^2}}\right)}\right].
\end{eqnarray}

The pairing potential in Eq.~(\ref{App-Gap-Eq-even-GE}) can be expanded as
\begin{eqnarray}
\label{App-Gap-Eq-even-V-expand}
V\left(\omega- \omega^{\prime}\right)+V\left(\omega+ \omega^{\prime}\right) \approx 2V(\omega)\theta\left(\omega-\omega^{\prime}\right) +2V(\omega^{\prime})\theta\left(\omega^{\prime}-\omega\right).
\end{eqnarray}
Then, the gap equation reads
\begin{eqnarray}
\label{App-Gap-Eq-even-Delta-1}
\Delta_{\rm even}(\omega) = 2\left[\int_0^{\omega}d\omega^{\prime} V(\omega) f_{\rm even}(\omega^{\prime}) +\int_{\omega}^{\infty}d\omega^{\prime} V(\omega^{\prime}) f_{\rm even}(\omega^{\prime})\right].
\end{eqnarray}
Unlike the case of the OF pairing considered in Sec.~II.A, the gap equation for the EF gap depends on the pairing potential itself.

In order to rewrite the integral equation in terms of the differential one, let us take the derivative with respect to frequency from both sides in Eq.~(\ref{App-Gap-Eq-even-Delta-1}), i.e.,
\begin{eqnarray}
\label{App-Gap-Eq-even-Delta-2}
\Delta^{\prime}_{\rm even}(\omega) = 2V^{\prime}(\omega) \int_0^{\omega}d\omega^{\prime} f_{\rm even}(\omega^{\prime}).
\end{eqnarray}
The second derivative reads
\begin{eqnarray}
\label{App-Gap-Eq-even-Delta-3}
\Delta_{\rm even}^{\prime \prime}(\omega) = 2\left[V^{\prime \prime}(\omega) \int_0^{\omega}d\omega^{\prime} f_{\rm even}(\omega^{\prime}) + V^{\prime}(\omega)f_{\rm even}(\omega)\right] = 2\left[ V^{\prime \prime}(\omega) \frac{\Delta_{\rm even}^{\prime}(\omega)}{2 V^{\prime}(\omega)} +V^{\prime}(\omega)f_{\rm even}(\omega)\right].
\end{eqnarray}
Finally, we obtain the following differential gap equation for the EF pairing:
\begin{eqnarray}
\label{App-Gap-Eq-even-Delta-GE}
\Delta_{\rm even}^{\prime \prime}(\omega) -\Delta_{\rm even}^{\prime}(\omega) V^{\prime \prime}(\omega)\left[V^{\prime}(\omega)\right]^{-1} =  2 V^{\prime}(\omega)f_{\rm even}(\omega)
\end{eqnarray}
with the boundary conditions
\begin{eqnarray}
\label{App-Gap-Eq-even-Delta-BC-1}
&&\Delta_{\rm even}^{\prime}(\omega) \left[V^{\prime}(\omega)\right]^{-1} \Big|_{\omega\to0}=0,\\
\label{App-Gap-Eq-even-Delta-BC-2}
&&\left\{\Delta_{\rm even}^{\prime}(\omega)- \Delta_{\rm even}(\omega) V^{\prime}(\omega) \left[V(\omega)\right]^{-1} \right\}\Big|_{\omega\to\infty}=0.
\end{eqnarray}
The latter are obtained from Eq.~(\ref{App-Gap-Eq-even-Delta-2}) at $\omega\to0$ as well as Eqs.~(\ref{App-Gap-Eq-even-Delta-1}) and (\ref{App-Gap-Eq-even-Delta-2}) at $\omega\to\infty$.

Finally, by using Eq.~(\ref{App-Gap-Eq-even-Delta-2}), it is straightforward to obtain the equation for the potential, i.e.,
\begin{eqnarray}
\label{App-Gap-Eq-even-V-eq}
V^{\prime}(\omega) = \Delta^{\prime}_{\rm even}(\omega)\left[2\int_0^{\omega}d\omega^{\prime} f_{\rm even}(\omega^{\prime}) \right]^{-1}.
\end{eqnarray}
The corresponding boundary condition follows from Eq.~(\ref{App-Gap-Eq-even-Delta-2}) in the limit $\omega\to\infty$, i.e.,
\begin{eqnarray}
\label{App-Gap-Eq-even-V-BC}
V(\omega)\Big|_{\omega\to\infty} = \Delta_{\rm even}(\omega) \left[2\int_0^{\omega}d\omega^{\prime} f_{\rm even}(\omega^{\prime}) \right]^{-1}\Big|_{\omega\to\infty}.
\end{eqnarray}
As expected, the pairing potential is completely fixed for the EF gaps.

\section{IV. Derivatives from the pairing potentials}
\label{sec:App-derivatives}

In this short section, the derivatives with respect to frequency from the OF pairing potential are presented for an ansatz considered in the min text, i.e.,
\begin{eqnarray}
\label{App-Wick-Delta-2-new}
\Delta_{\rm odd}(\omega)= \alpha \frac{\Lambda_k}{\omega},
\end{eqnarray}
which is considered in the main text. The dependence of the gap as well as the potential derivative on frequency is shown in Fig.~\ref{fig:App-Gap-Eq-inter-SS-swave-Delta-V-prime}a). It is clear that the corresponding derivative quickly diminishes with the frequency at large $\omega$ but diverges at $\omega\to0$. As for the dependence of the gap on the derivative strength, the corresponding results at a few values of the frequency are shown in Fig.~\ref{fig:App-Gap-Eq-inter-SS-swave-Delta-V-prime}b). As one can see, there is a critical value of the potential derivative $V^{\prime}(\omega)$ strength that corresponds to the generation of the gap. Such a value tends to diminish with the frequency, which can be easily understood since the gap also decreases with $\omega$. Therefore, the higher the frequency, the lower the OF gap and the smaller potential derivative is needed to generate it.

\begin{figure*}[!ht]
\begin{center}
\includegraphics[height=0.4\textwidth]{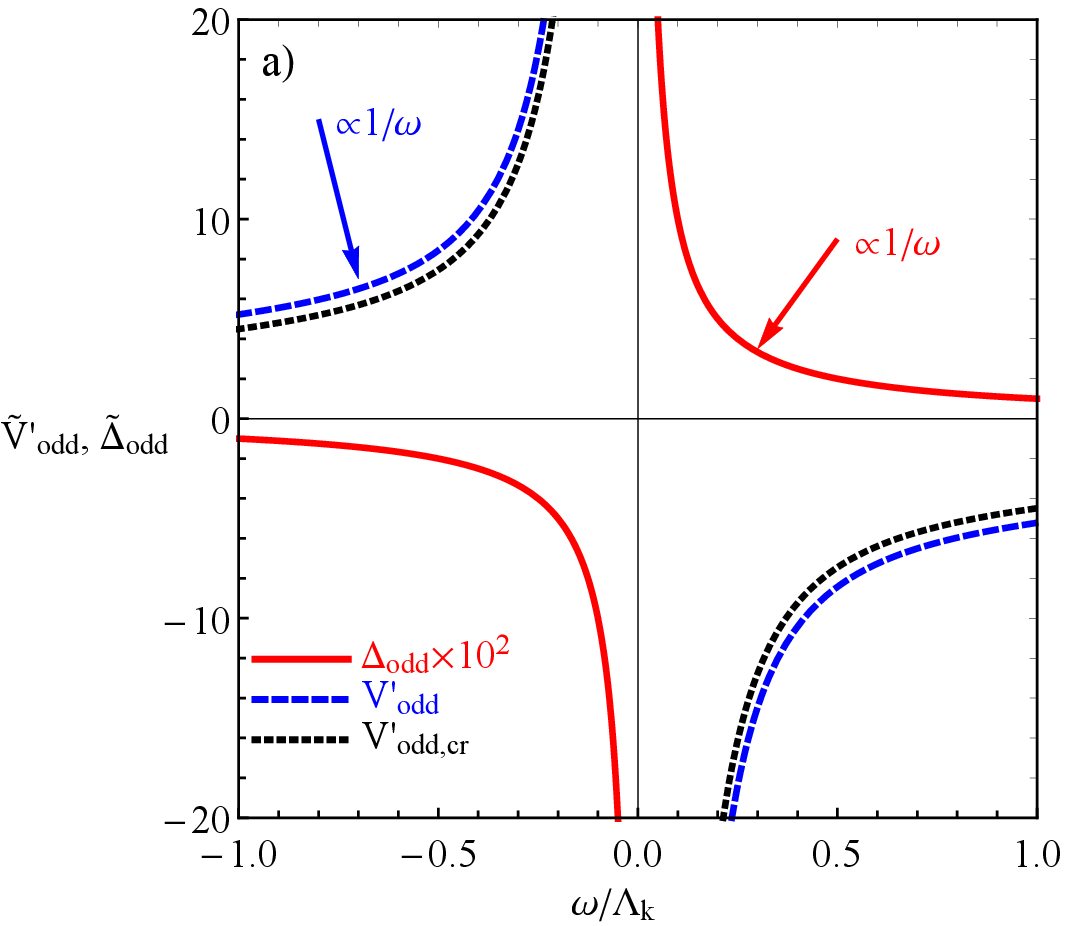}\hfill
\includegraphics[height=0.4\textwidth]{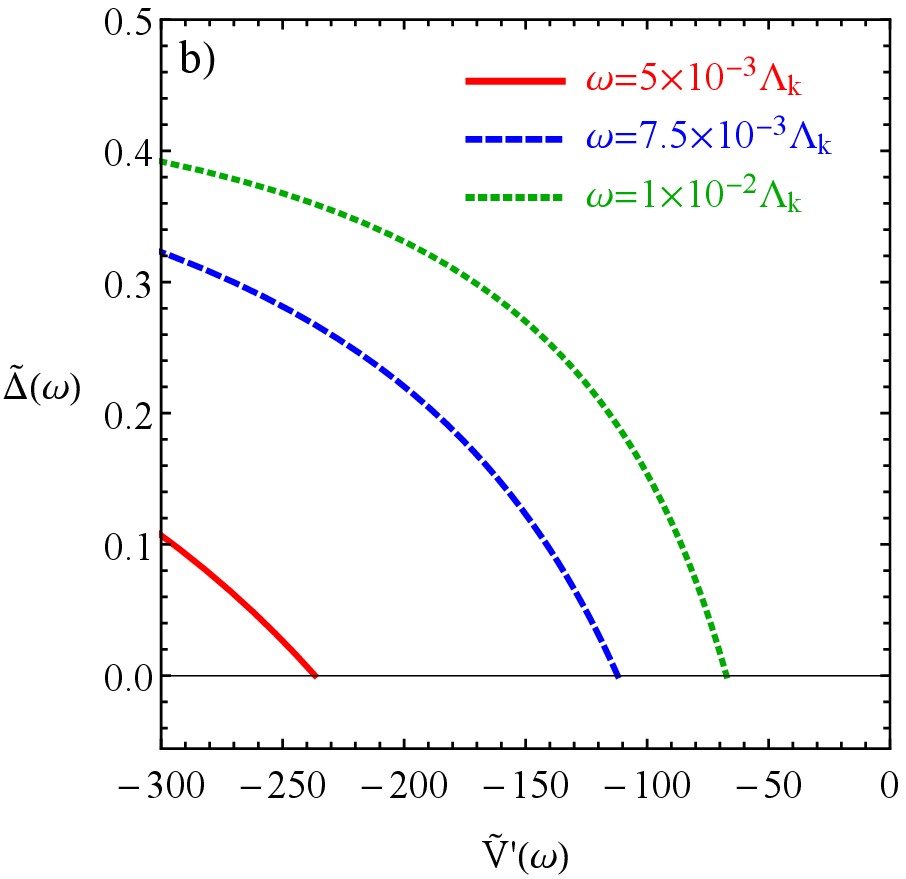}
\end{center}
\caption{Panel a): The derivative of the pairing potential with respect to frequency as functions of $\omega$ at $\alpha=10^{-2}\Lambda_k$. Red solid line corresponds to the gap given in Eq.~(\ref{App-Wick-Delta-2-new}) and the blue dashed line is the derivative with respect to frequency from the corresponding potential $V^{\prime}(\omega)$.
Panel b): The dependence of the OF gap on the frequency derivative from the pairing potential. The frequency is $\omega=5\times10^{-3}\Lambda_k$ (red solid line), $\omega=7.5\times10^{-3}\Lambda_k$ (blue dashed line), and $\omega=1\times10^{-2}\Lambda_k$ (green dotted line).
In both panels, $\tilde{\Delta}(\omega)=\Delta(\omega)/\Lambda_k$ and $\tilde{V}^{\prime}(\omega)=V(\omega)/(\Lambda_k|V_{\rm crit}|)$, where $V_{\rm crit} = -8\pi^2 v_F^3/\Lambda_k^2$. Note also that we used a finite frequency cutoff $\Lambda_{\omega}=2\Lambda_k$ in the numerical calculations. We checked, however, that the results are almost insensitive to the value of this cutoff.
}
\label{fig:App-Gap-Eq-inter-SS-swave-Delta-V-prime}
\end{figure*}

\section{V. Spectral functions and density of states}
\label{sec:App-observables}

In this section, we discuss a few possible observables such as the spectral function $A(\omega,\mathbf{k})$ as well as the electron density of states $\nu(\omega)$. Let us start with the spectral function, which is defined as
\begin{eqnarray}
\label{App-Observables-AG}
A(\omega,\mathbf{k}) &=& \frac{1}{2\pi i} \left[G(\omega-i0,\mathbf{k})-G(\omega+i0,\mathbf{k})\right]_{\mu=0},
\end{eqnarray}
where $G(\omega\pm i0)$ are the retarded $G^{\rm R}(\omega,\mathbf{k})$ and advanced $G^{\rm A}(\omega,\mathbf{k})$ Green's functions, respectively. Then, the Green's function can be obtained as
\begin{eqnarray}
\label{App-Observables-G-AG}
G(\Omega,\mathbf{k}) = \int \frac{d\omega A(\omega,\mathbf{k})}{\Omega+\mu-\omega}.
\end{eqnarray}

By using Green's functions defined in Eqs.~(\ref{App-OP-Weyl-Green-G-intra-SS-swave-even}) and (\ref{App-OP-Weyl-Green-G-inter-SS-swave}), we derive the following spectral functions:
\begin{eqnarray}
\label{App-Observables-AG-even}
A_{\rm even}(\omega,\mathbf{k}) &=& \sign{\omega} \left[\omega \mathds{1}_4 +v_F \gamma^0 (\bm{\gamma}\cdot\mathbf{k})\right] \delta\left(\omega^2-|\Delta(\omega)|^2 -v_F^2k^2\right),\\
\label{App-Observables-AG-odd}
A_{\rm odd}(\omega,\mathbf{k}) &=& \frac{\sign{\omega}}{4v_F k |\Delta(\omega)|} \left\{\omega \mathds{1}_4 \left[\omega^2 -v_F^2k^2 -|\Delta(\omega)|^2\right] +v_F \gamma^0 (\bm{\gamma}\cdot\mathbf{k}) \left[\omega^2 -v_F^2k^2 +|\Delta(\omega)|^2\right]\right\} \nonumber\\
&\times&\left\{\delta\left[\omega^2-\left(|\Delta(\omega)| +v_Fk\right)^2\right] -\delta\left[\omega^2-\left(|\Delta(\omega)| -v_Fk\right)^2\right] \right\}.
\end{eqnarray}

The other important observable is the electron DOS, which is defined as
\begin{eqnarray}
\label{App-Observables-DOS-def}
\nu(\omega) = -\frac{1}{\pi} \mbox{Im} \int \frac{d^3k}{(2\pi)^3} \mbox{tr}\left[G^{\rm R}(\omega,\mathbf{k})\right].
\end{eqnarray}
By using Green's functions given in Eqs.~(\ref{App-OP-Weyl-Green-G-intra-SS-swave-even}) and (\ref{App-OP-Weyl-Green-G-inter-SS-swave}), we obtain the following electron DOS in the cases of the even- and odd-frequency pairings:
\begin{eqnarray}
\label{App-Observables-DOS-even}
\nu^{\rm even}(\omega) &=& \frac{\sign{\omega}}{2\pi^2} \int_0^{\Lambda_k/v_F} \frac{k\,dk}{v_F \mu} \left\{(\omega-\mu)\left[(\omega+\mu)^2 - |\Delta(\omega)|^2 -v_F^2k^2\right] -2\mu |\Delta(\omega)|^2\right\} \theta(\omega^2-|\Delta(\omega)|^2) \nonumber\\
&\times&\left\{\delta\left[\omega^2 - |\Delta(\omega)|^2 -\left(\mu +v_Fk\right)^2\right] -\delta\left[\omega^2 - |\Delta(\omega)|^2 -\left(\mu -v_Fk\right)^2\right] \right\}
\end{eqnarray}
and
\begin{eqnarray}
\label{App-Observables-DOS-odd}
\nu^{\rm odd}(\omega) &=& \frac{\sign{\omega}}{2\pi^2} \int_0^{\Lambda_k/v_F} \frac{k\,dk}{v_F \sqrt{|\Delta(\omega)|^2+\mu^2}} \left\{(\omega-\mu)\left[(\omega+\mu)^2 - |\Delta(\omega)|^2 -v_F^2k^2\right] -2\mu |\Delta(\omega)|^2\right\} \nonumber\\
&\times&\left\{\delta\left[\omega^2 -\left(v_Fk+\sqrt{\mu^2 +|\Delta(\omega)|^2}\right)^2\right] -\delta\left[\omega^2 -\left(v_Fk-\sqrt{\mu^2 +|\Delta(\omega)|^2}\right)^2\right] \right\},
\end{eqnarray}
respectively. The integrals over momentum in Eqs.~(\ref{App-Observables-DOS-even}) and (\ref{App-Observables-DOS-odd}) can be straightforwardly calculated, which leads, however, to cumbersome expressions. Therefore, we do not present them here.

Let us investigate the evolution of the spectral weight during the gap opening.
To start with, we present the trace of the spectral function $A_{\rm tr}=\mbox{tr}\,A(\omega,\mathbf{k})$ and the corresponding electron DOS for the even-frequency gap $\Delta(\omega)=\alpha$ in Figs.~\ref{fig:App-Observables-DOS-even}a) and \ref{fig:App-Observables-DOS-even}b), respectively. As expected, the even-frequency gap pushes the states away from the region of small $\omega$. The corresponding spectral weight is recovered at the cutoff as can be seen in Fig.~\ref{fig:App-Observables-DOS-even}b). As in the case of 2D Dirac semimetals (for the corresponding discussion, see, e.g., Ref.~[S7]), no coherence peaks appear at the charge neutrality point.
Note that, in order to obtain the results for the spectral function, the $\delta$-functions in Eqs.~(\ref{App-Observables-AG-even}) and (\ref{App-Observables-AG-odd}) were broadened via the Lorentzian distribution, i.e.,
\begin{eqnarray}
\label{App-Observables-delta-Gamma}
\delta(x)\to \frac{1}{\pi} \frac{\Gamma}{x^2+\Gamma^2}
\end{eqnarray}
with $\Gamma=10^{-4}\Lambda_{k}$.

\begin{figure*}[!ht]
\begin{center}
\includegraphics[height=0.4\textwidth]{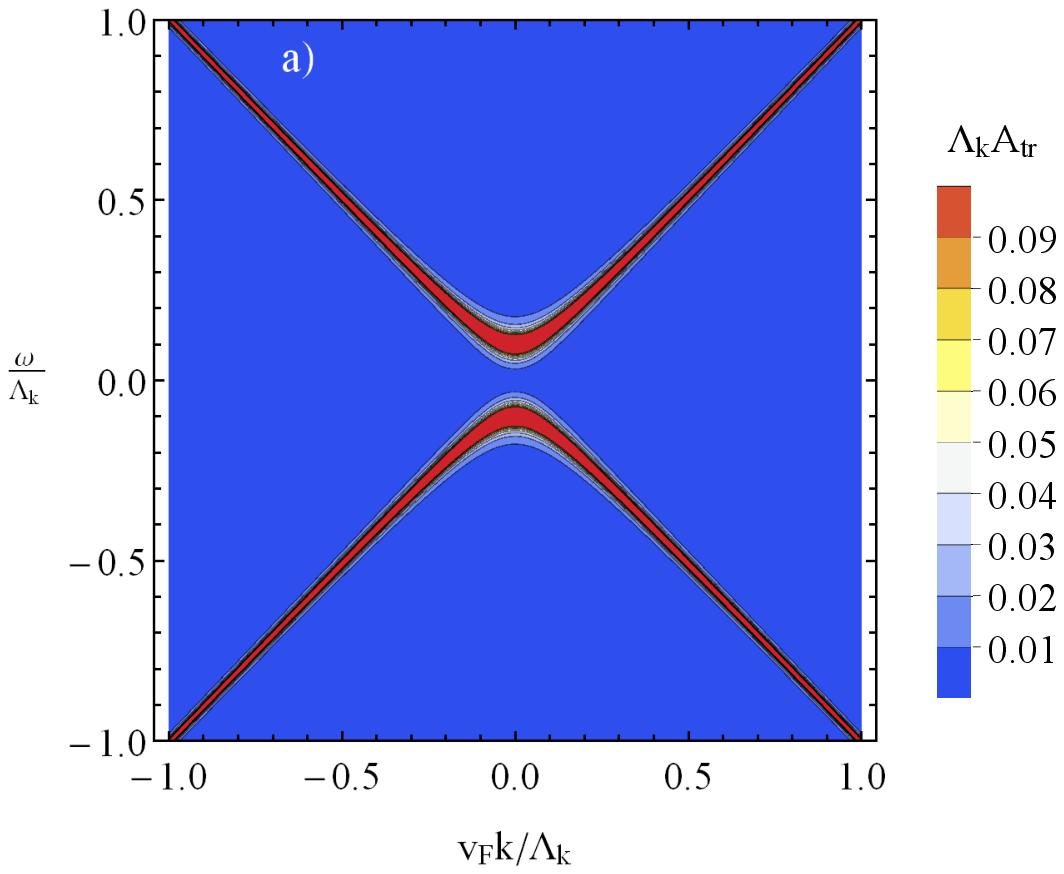}\hfill
\includegraphics[height=0.4\textwidth]{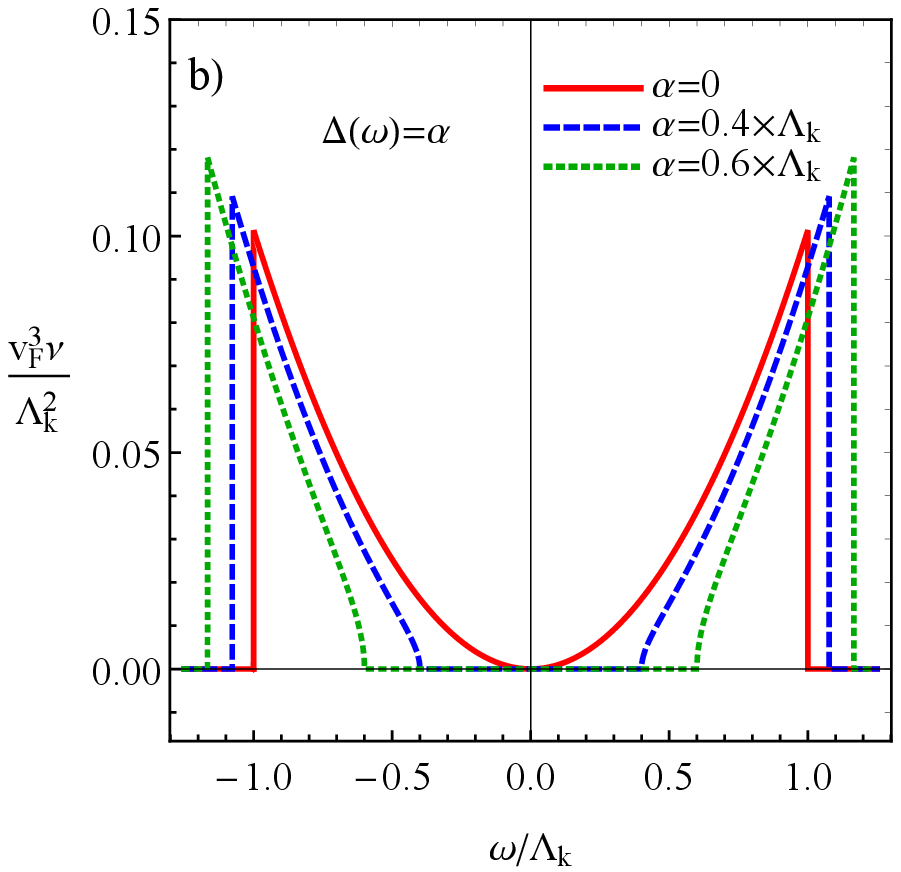}
\end{center}
\caption{The dependence of the trace of the spectral function $A_{\rm tr}=\mbox{tr}\,A(\omega,\mathbf{k})$ on the frequency $\omega$ and the momentum $\mathbf{k}$ (panel a)) and the electron DOS $\nu(\omega)$ on the frequency $\omega$ (panel b)) for $\Delta(\omega)=\alpha$. We set $\mu=0$.
As expected, the formation of the gap pushes the states away from the region of small frequencies. The corresponding spectral weight is recovered at the cutoff $\omega\approx\Lambda_k$.
}
\label{fig:App-Observables-DOS-even}
\end{figure*}

The trace of the spectral function $A_{\rm tr}=\mbox{tr}\,A(\omega,\mathbf{k})$ and the corresponding electron DOS for the odd-frequency gap $\Delta(\omega)=\alpha\Lambda_k/\omega$ are shown in Figs.~\ref{fig:App-Observables-DOS-9}a) and \ref{fig:App-Observables-DOS-9}b), respectively. As one can see, the results for the odd-frequency gap are more complicated. In particular, there are four separated energy branches. For sufficiently large $\omega$, they merge together because the effect of the odd-frequency gap $\Delta=\alpha \Lambda_{k}/\omega$ diminishes with the frequency.
The results for the corresponding DOS in the case of the odd-frequency pairing are shown in Fig.~\ref{fig:App-Observables-DOS-9}b). As one can see, the DOS has a complicated structure with several cusp-like features. By comparing the results for the DOS and the trace of the spectral function given in Figs.~\ref{fig:App-Observables-DOS-9}a) and \ref{fig:App-Observables-DOS-9}b), respectively, we found that these cusps appear exactly at frequencies where $A_{\rm tr}$ at $k_z=\Lambda_k$ is nonvanishing. Indeed, the two cusps at $|\omega|\gtrsim0.5\Lambda_k$ correspond to the two splitted high-frequency branches.
The weakly separated features at small frequencies $|\omega|\lesssim 0.5\Lambda_k$, which are shown in detail in Fig.~3b) in the main text, stem from the almost flat low-frequency branches in $A_{\rm tr}$.
In addition, we note that, from the technical viewpoint, the appearance of the cusps is directly related to the finite momentum cutoff $\Lambda_k$. We checked that the DOS for the odd-frequency state shows a single peak at $\omega\to0$ for $\Lambda_k\to\infty$.

\begin{figure*}[!ht]
\begin{center}
\includegraphics[height=0.4\textwidth]{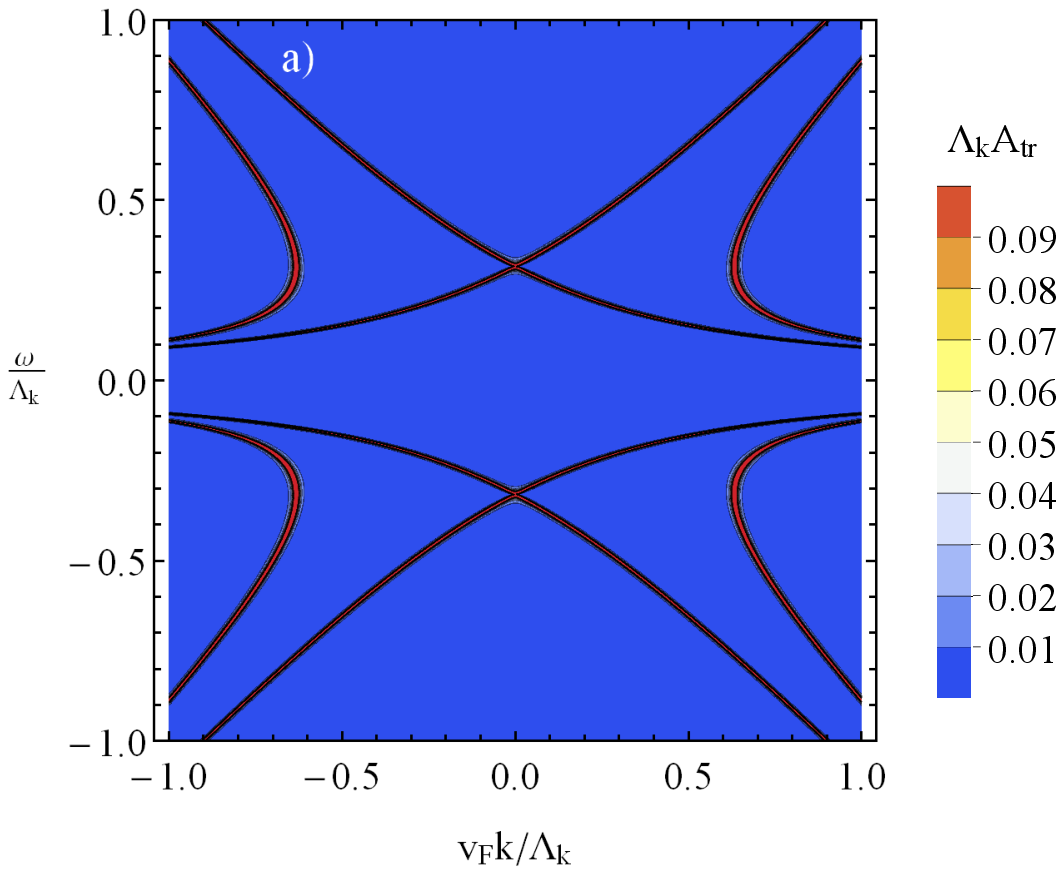}\hfill
\includegraphics[height=0.4\textwidth]{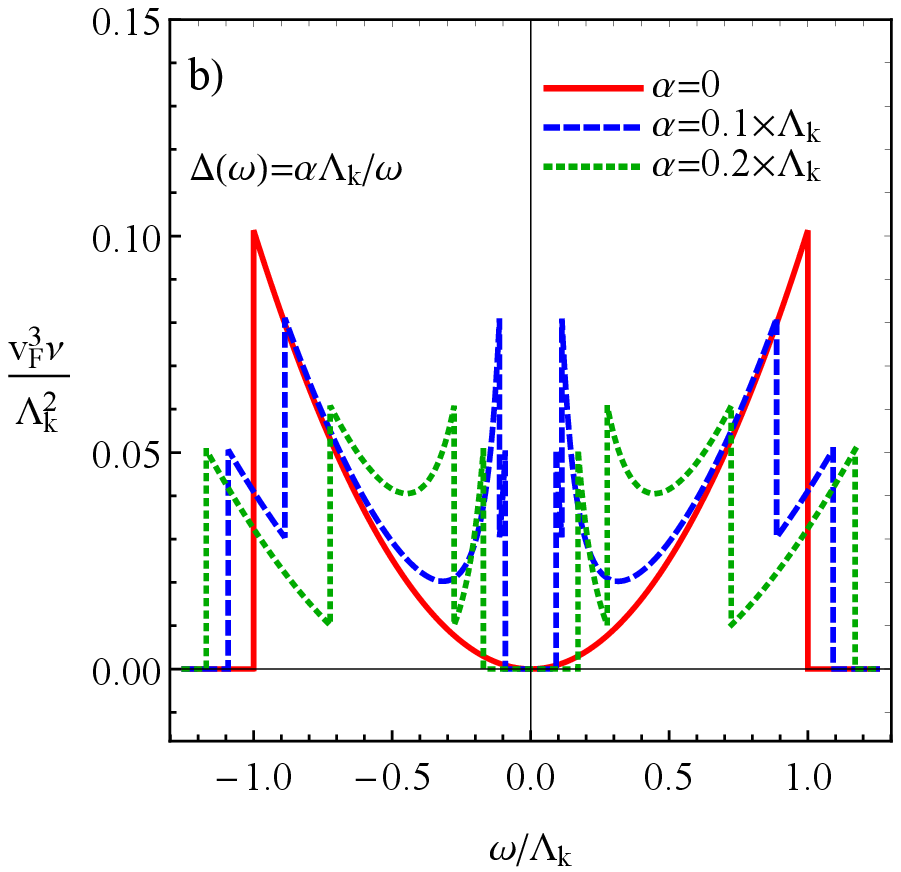}
\end{center}
\caption{The dependence of the trace of the spectral function $A_{\rm tr}=\mbox{tr}\,A(\omega,\mathbf{k})$ on the frequency $\omega$ and the momentum $\mathbf{k}$ (panel a)) and the electron DOS $\nu(\omega)$ on the frequency $\omega$ (panel b)) for $\Delta(\omega)=\alpha\Lambda_k/\omega$. We set $\mu=0$.
}
\label{fig:App-Observables-DOS-9}
\end{figure*}

In order to demonstrate the universality of some features, let us present $A_{\rm tr}$ and the corresponding electron DOS for two more possible ansatzes of odd-frequency gaps. In particular, the results for $\Delta=\alpha \omega/\sqrt{\omega^2+\beta^2\Lambda_k^2}$ and $\Delta=\alpha \omega \Lambda_k/\left(\omega^2+\beta^2\Lambda_k^2\right)$ are presented in Figs.~\ref{fig:App-Observables-DOS-4} and \ref{fig:App-Observables-DOS-2}. Comparing the results in Figs.~\ref{fig:App-Observables-DOS-9}, \ref{fig:App-Observables-DOS-4}, and \ref{fig:App-Observables-DOS-2}, one can see that the splitting of the energy branches (poles of the spectral function) is a universal feature of the odd-frequency pairing. The details of the splitting depend, however, on the gap ansatz. We suggest that the experimental observation of the splitting of the spectral function branches could provide an evidence in favor of the spin-singlet, s-wave, Berezinskii pairing in 3D Dirac semimetals.

\begin{figure*}[!ht]
\begin{center}
\includegraphics[height=0.4\textwidth]{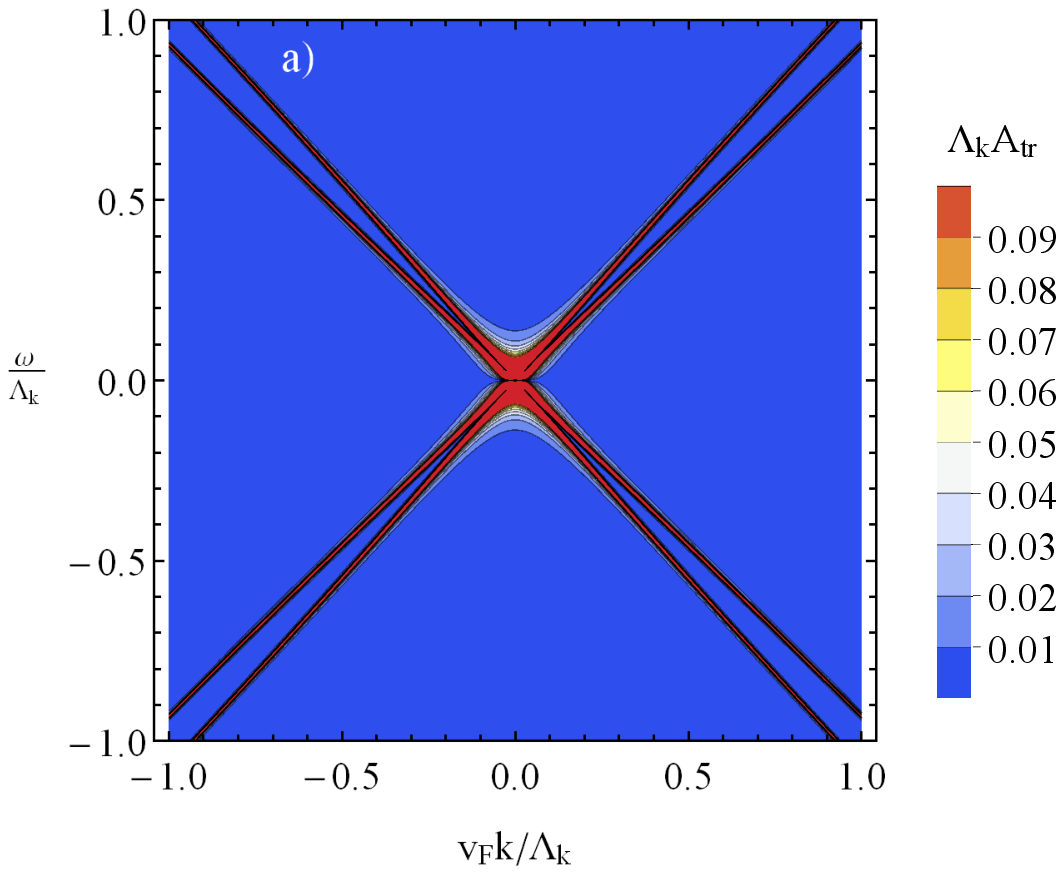}\hfill
\includegraphics[height=0.4\textwidth]{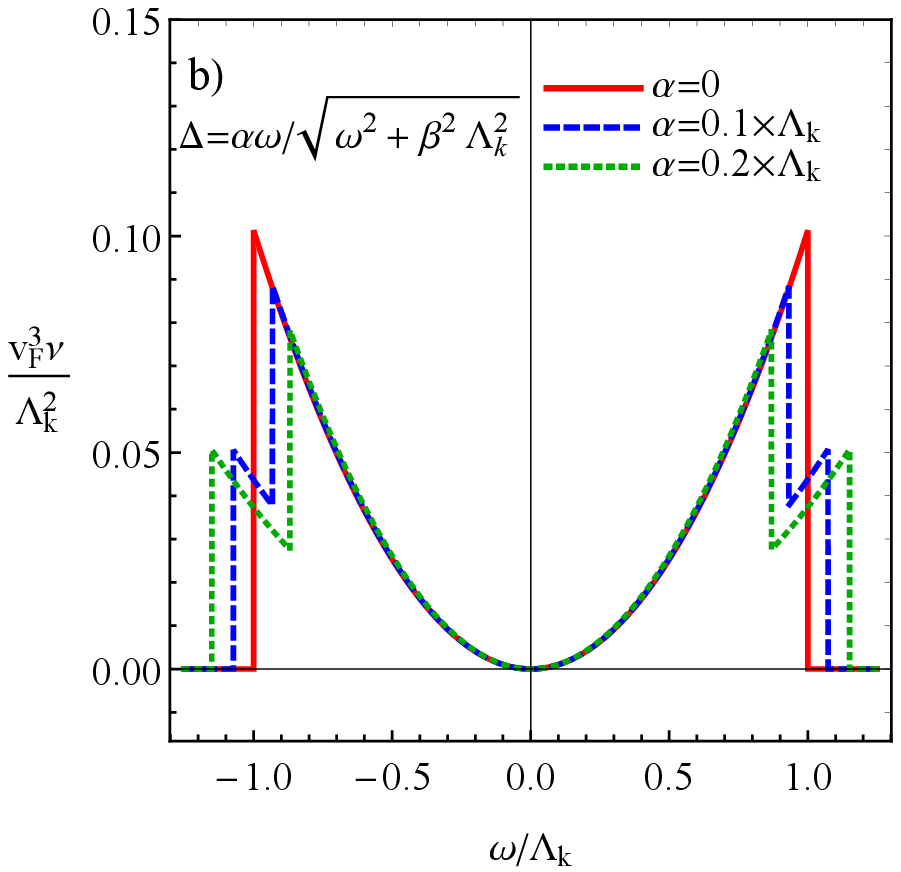}
\end{center}
\caption{The dependence of the trace of the spectral function $A_{\rm tr}=\mbox{tr}\,A(\omega,\mathbf{k})$ on the frequency $\omega$ and the momentum $\mathbf{k}$ (panel a)) and the electron DOS $\nu(\omega)$ on the frequency $\omega$ (panel b)) for $\Delta=\alpha \omega/\sqrt{\omega^2+\beta^2\Lambda_k^2}$. We set $\mu=0$ and $\beta=1$.
}
\label{fig:App-Observables-DOS-4}
\end{figure*}

\begin{figure*}[!ht]
\begin{center}
\includegraphics[height=0.4\textwidth]{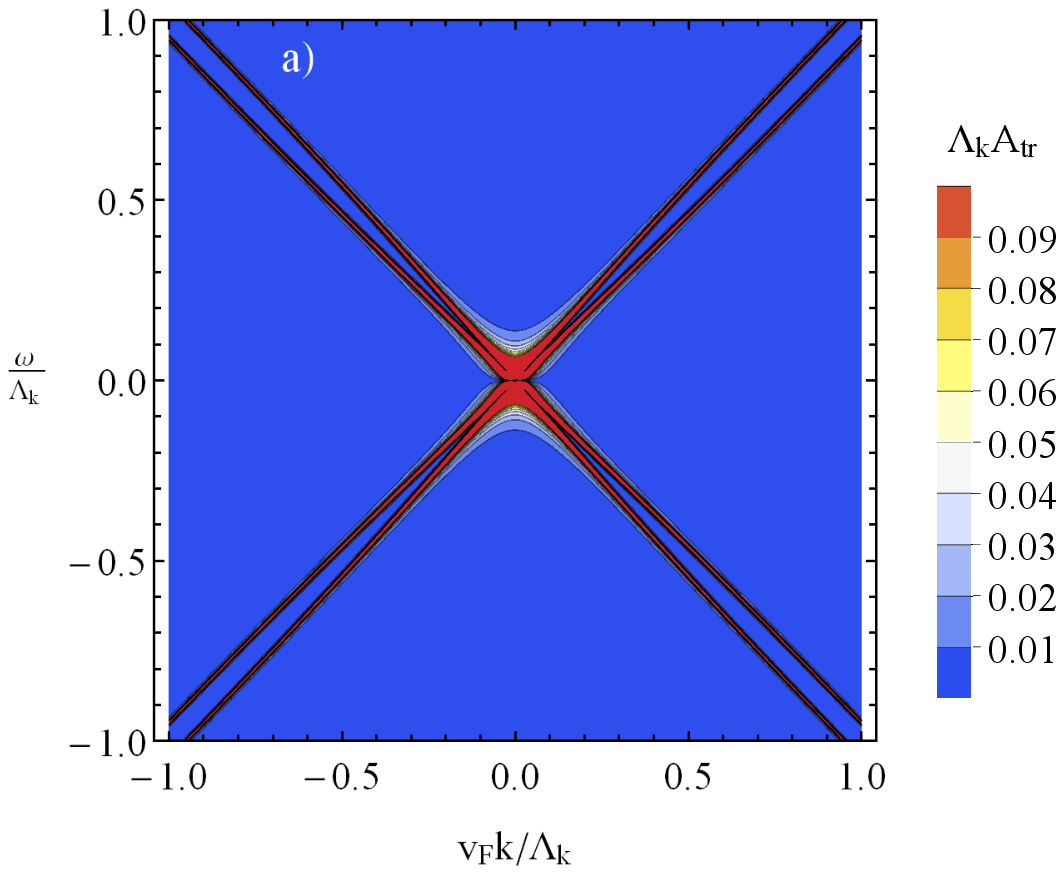}\hfill
\includegraphics[height=0.4\textwidth]{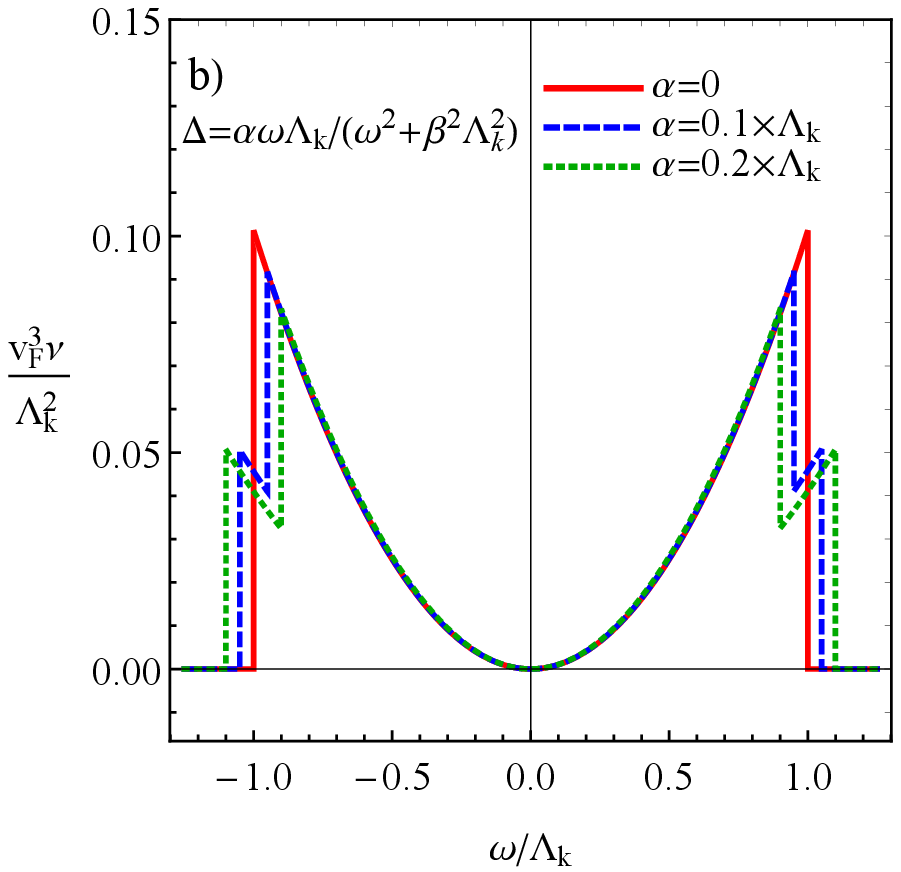}
\end{center}
\caption{The dependence of the trace of the spectral function $A_{\rm tr}=\mbox{tr}\,A(\omega,\mathbf{k})$ on the frequency $\omega$ and the momentum $\mathbf{k}$ (panel a)) and the electron DOS $\nu(\omega)$ on the frequency $\omega$ (panel b)) for $\Delta=\alpha \omega \Lambda_k/\left(\omega^2+\beta^2\Lambda_k^2\right)$. We set $\mu=0$ and $\beta=1$.
}
\label{fig:App-Observables-DOS-2}
\end{figure*}

Next, let us discuss the evolution of the electron DOS at nonzero electric chemical potential $\mu$. The corresponding results at a few values of $\mu$ in the case of the even- and odd-frequency pairings are shown in Figs.~\ref{fig:App-Observables-DOS-mu}a) and \ref{fig:App-Observables-DOS-mu}b), respectively. The even-frequency gap pushes the states away from the region of small frequencies. As in the case of 2D Dirac semimetals (for the corresponding discussion, see Fig.~\ref{fig:App-Observables-DOS-mu-2D} and, e.g., Ref.~[S8]), while no coherence peaks appear at the charge neutrality point, they are generically present at $\mu\neq0$. On the other hand, the formation of the coherence peaks is not necessary the case for the odd-frequency gap.

\begin{figure*}[!ht]
\begin{center}
\includegraphics[height=0.4\textwidth]{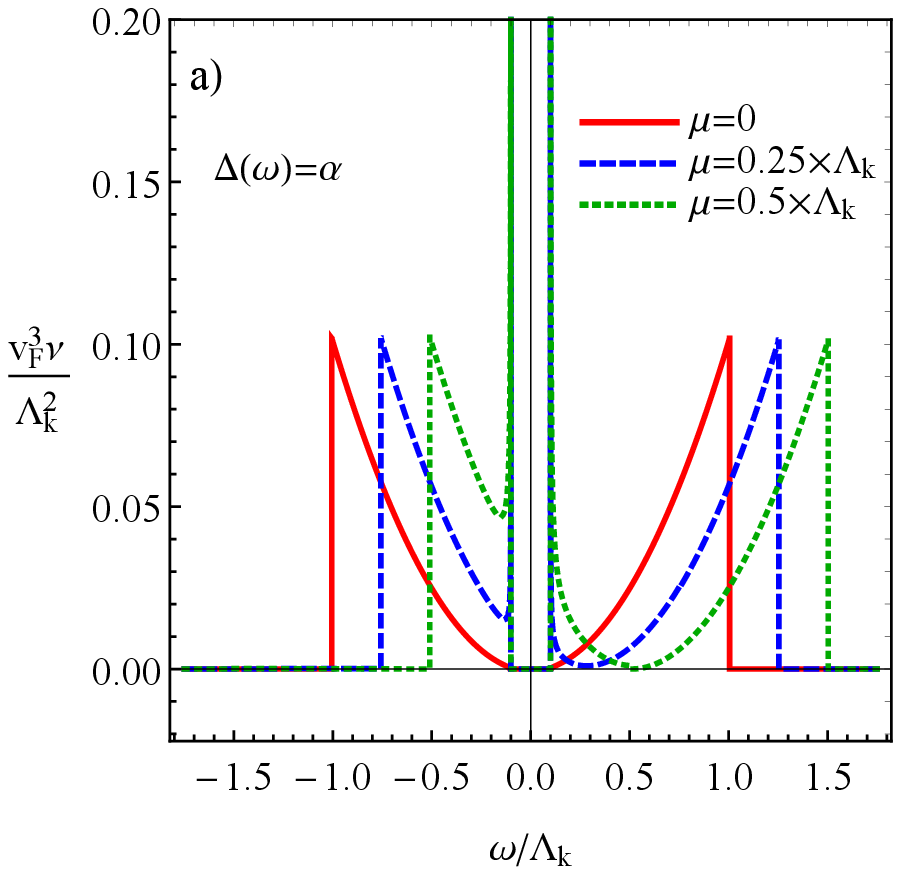}\hfill
\includegraphics[height=0.4\textwidth]{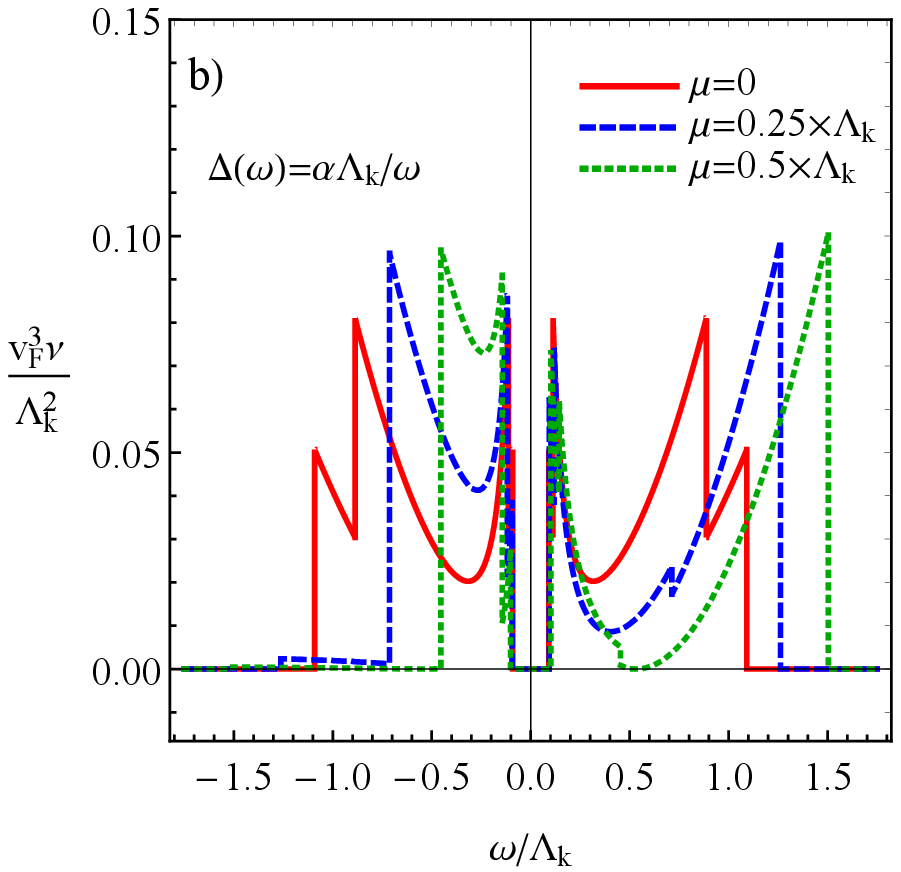}
\end{center}
\caption{The dependence of the electron DOS $\nu(\omega)$ on the frequency $\omega$ for $\Delta=\alpha$ (panel a)) and $\Delta=\alpha \Lambda_k/\omega$ (panel b)) at a few values of the electric chemical potential $\mu$. We set $\alpha=0.1\,\Lambda_k$.
}
\label{fig:App-Observables-DOS-mu}
\end{figure*}

It is reasonable to suggest that the intersections in the case of the odd-frequency gap shown in Fig.~\ref{fig:App-Observables-DOS-9}a), might have a nontrivial topology. In order to clarify this, let us calculate the following topological invariant~[S8,S9]:
\begin{eqnarray}
\label{App-Observables-N-gamma}
N^{(\gamma)} = \int dS^{\rho}\Omega^{(\gamma)}_{\rho},
\end{eqnarray}
where
\begin{eqnarray}
\label{App-Observables-Omega-gamma}
\Omega^{(\gamma)}_{\rho} = -\frac{1}{24\pi^2} \epsilon_{\mu \nu \lambda \rho} \mbox{tr}\left[\gamma G_{\rm N} \partial_{k_{\mu}}G_{\rm N}^{-1} G_{\rm N} \partial_{k_{\nu}}G_{\rm N}^{-1} G_{\rm N} \partial_{k_{\lambda}}G_{\rm N}^{-1}\right],
\end{eqnarray}
is the 4D analogue of the conventional Berry curvature, $\gamma$ is $8\times8$ matrix that either commutes or anticommutes with the Green's function, $k_{\mu}=\left(ik_0,-v_F\mathbf{k}\right)$ is the momentum 4-vector with $\omega\to i k_0$, and $\int dS^{\rho}$ is performed over the 3D sphere embracing singular points. It is straightforward to verify for usual Weyl semimetals that the vector field $\left(\Omega^{(\gamma)}_{4}, \Omega^{(\gamma)}_{3} \right)$ in the $\left(k_0,k_z\right)$ space at $\gamma=\mathds{1}_8$ has the same monopole-like structure as the Berry curvature in momentum space. We checked, however, that the field lines of $\Omega^{(\gamma)}_{\rho}$, which is defined in Eq.~(\ref{App-Observables-Omega-gamma}), are trivial for Dirac semimetals both with even- and odd-frequency gaps considered in this study. Thus, in this sense, the crossings in Fig.~\ref{fig:App-Observables-DOS-9}a) are topologically trivial. Clearly, the topological signatures of the odd-frequency pairing deserve further investigation.

To conclude, let us summarize the key features of the spectral weight in the case of even- and odd-frequency pairings:
\begin{itemize}
  \item Even-frequency (BCS) pairing
  \begin{itemize}
    \item The gap pushes the states away from the region of small $\omega$. The corresponding spectral weight is recovered at the cutoff.
    \item No coherence peaks are formed at the charge neutrality point $\mu=0$. The peaks, however, appear for $\mu\neq0$.
  \end{itemize}
  \item Odd-frequency pairing
    \begin{itemize}
    \item There is a redistribution of the spectral weight due to the formation of the cusp-like features at the cutoff.
    \item The coherence peaks are generically absent even at $\mu\neq0$.
  \end{itemize}
\end{itemize}

\section{VI. 2D Dirac semimetals}
\label{sec:App-2D}

In this section, we discuss the case of 2D Dirac semimetals at the charge neutrality point. We employ the same Hamiltonian given by Eq.~(\ref{App-model-free-Hamiltonian-Weyl-rel}), where $k_z=0$. Physically, this model corresponds to graphene, where the chirality is equal to the valley index and (pseudo)spin corresponds to the sublattice degree of freedom. Then, in order to study the effects of dimensionality on the superconducting pairing and spectroscopic response, we can consider the same even- and odd-frequency gaps given by Eqs.~(\ref{App-model-OF-gaps-intra-SS-swave}) and (\ref{App-model-OF-gaps-inter-SS-swave}), respectively. As is easy to verify, the Green functions will be given by the same expressions as in 3D case albeit with $k_z=0$ (see Sec.~\ref{sec:App-model-OF-EA}). The derivation of the gap equations and equations for the pairing potentials is identical to that in Sec.~III, where, however, functions $\hat{f}_{\rm odd}(\omega^{\prime})$  and $\hat{f}_{\rm even}(\omega^{\prime})$ are given by different expressions. Their explicit form reads
\begin{eqnarray}
\label{App-Wick-f-odd-def-2D}
\hat{f}_{\rm odd}(\omega^{\prime}) &=& \int \frac{d^2k}{(2\pi)^3} \hat{F}_{\rm odd}(\omega^{\prime}) = \gamma^0\gamma_5\frac{\Delta_{\rm odd}(\omega^{\prime})}{(2\pi)^2} \int_{0}^{\Lambda_k/v_F} kdk \frac{v_F^2k^2- (\omega^{\prime})^2-|\Delta_{\rm odd}(\omega^{\prime})|^2}{\left[(\omega^{\prime})^2+\left(v_Fk+|\Delta_{\rm odd}(\omega^{\prime})|\right)^2\right]\left[(\omega^{\prime})^2+\left(v_Fk-|\Delta_{\rm odd}(\omega^{\prime})|\right)^2\right]} \nonumber\\
&=& \gamma^0\gamma_5\frac{\sign{\Delta_{\rm odd}(\omega^{\prime})}}{(4\pi v_F)^2}  \Bigg\{ \Delta(\omega^{\prime}) \ln{\left(\frac{(\omega^{\prime})^2 +\left[|\Delta(\omega^{\prime})| +\Lambda_k\right]^2}{(\omega^{\prime})^2 +|\Delta(\omega^{\prime})|^2} \frac{(\omega^{\prime})^2 +\left[|\Delta(\omega^{\prime})| -\Lambda_k\right]^2}{(\omega^{\prime})^2 +|\Delta(\omega^{\prime})|^2}\right)} \nonumber\\
&+&2\omega^{\prime} \left[\arctan{\left(\frac{|\Delta(\omega^{\prime})|+\Lambda_k}{\omega^{\prime}}\right)} +\arctan{\left(\frac{|\Delta(\omega^{\prime})|-\Lambda_k}{\omega^{\prime}}\right)} -2\arctan{\left(\frac{|\Delta(\omega^{\prime})|}{\omega^{\prime}}\right)} \right]
\Bigg\},
\end{eqnarray}
and
\begin{eqnarray}
\label{App-Wick-f-even-def-2D}
\hat{f}_{\rm even}(\omega^{\prime}) &=& \int \frac{d^2k}{(2\pi)^3} \hat{F}_{\rm even}(\omega^{\prime}) = -\mathds{1}_4 \frac{\Delta_{\rm even}(\omega^{\prime})}{(2\pi)^2} \int_{0}^{\Lambda_k/v_F} kdk \frac{1}{(\omega^{\prime})^2+|\Delta_{\rm even}(\omega^{\prime})|^2+v_F^2k^2} \nonumber\\
&=& -\mathds{1}_4 \frac{\Delta_{\rm even}(\omega^{\prime})}{2(2\pi v_F)^2} \ln{\left(\frac{(\omega^{\prime})^2 +\Lambda_k^2 +|\Delta(\omega^{\prime})|^2}{(\omega^{\prime})^2 +|\Delta(\omega^{\prime})|^2}\right)}.
\end{eqnarray}

The results for the pairing potential and the superconducting gaps are presented in Figs.~\ref{fig:App-Gap-Eq-inter-SS-swave-Delta-2D}a) and \ref{fig:App-Gap-Eq-inter-SS-swave-Delta-2D}b), respectively. We used the same gap asatzes as in the 3D case considered the main text, i.e., $\Delta_{\rm even}(\omega)= \alpha$ and $\Delta_{\rm odd}(\omega)= \alpha \Lambda_k/\omega$. Note that, because of a different DOS, the critical potential (coupling) for the EF case is different, i.e.,
\begin{eqnarray}
\label{App-Delta-0-crit-2D}
V_{\rm crit} =g_{\rm cr}^{\rm even}= -\frac{4\pi v_F^2}{\Lambda_k}.
\end{eqnarray}
As can see by comparing the results in Fig.~2 in the main text and Fig.~\ref{fig:App-Gap-Eq-inter-SS-swave-Delta-2D}, the difference between the pairing potentials and gaps in 2D and 3D is quantitative rather than qualitative.

\begin{figure*}[!ht]
\begin{center}
\includegraphics[height=0.4\textwidth]{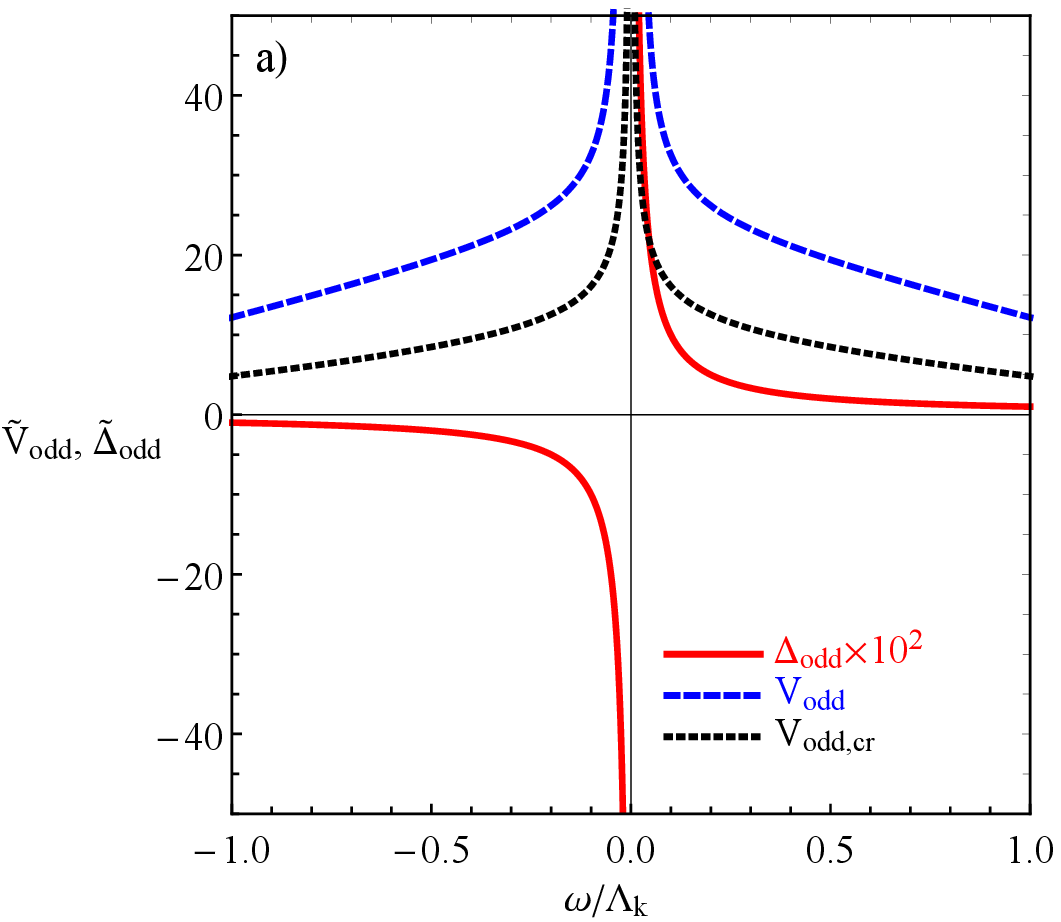}\hfill
\includegraphics[height=0.4\textwidth]{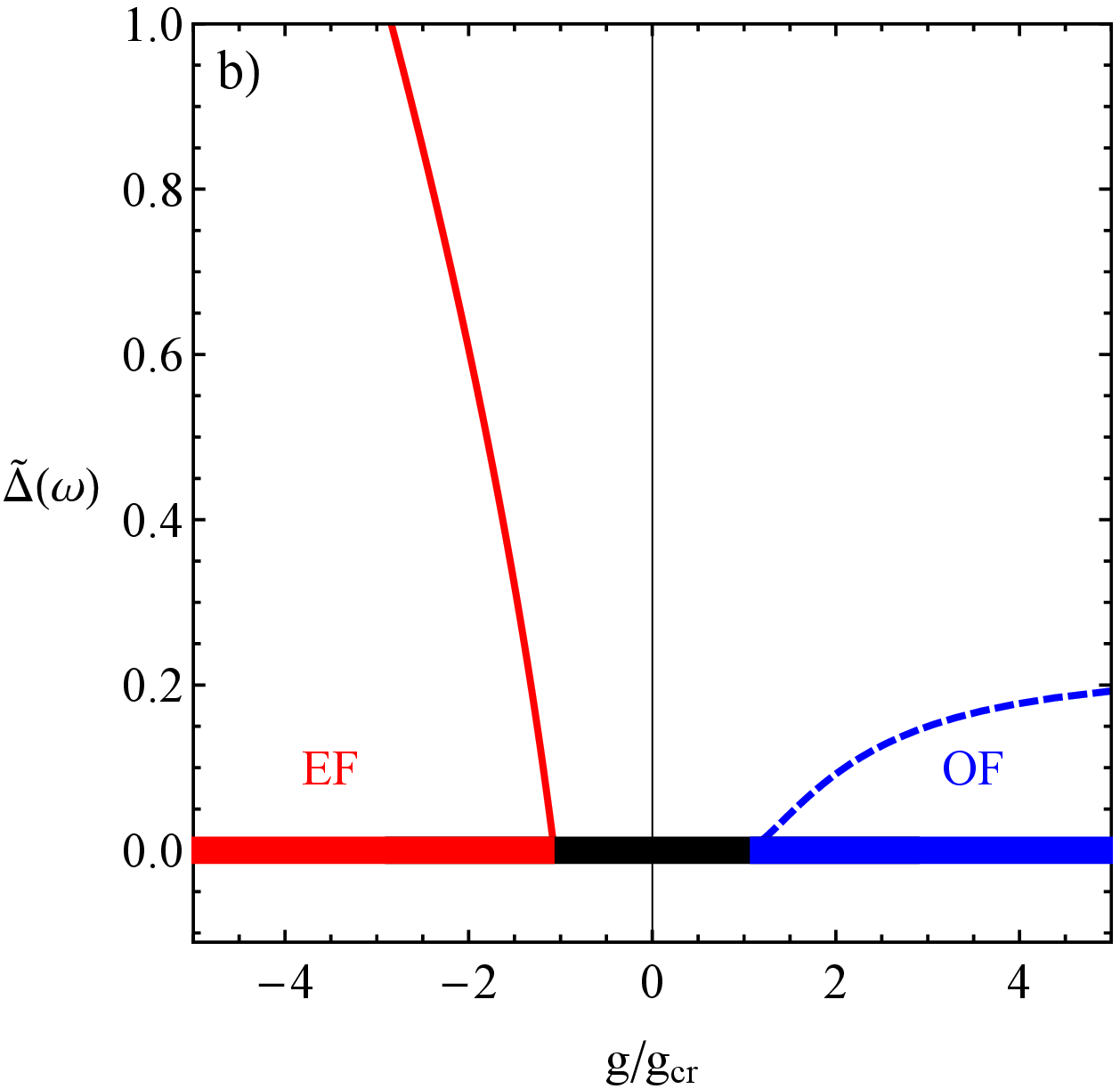}
\end{center}
\caption{Panel a): The odd-frequency gap and the corresponding pairing potential as functions of frequency at $\alpha=10^{-2}\Lambda_k$. Red solid lines correspond to the OF gap and blue dashed ones represent the corresponding potential. The critical value of the pairing potential obtained at $\alpha\to0$ is represented by a black dotted line. Panel b): The dependence of the gaps on the coupling constant for the EF (red solid line) and OF (blue dashed line) gaps.
The 1D phase diagram is shown by the thick red, blue, and black lines at $\tilde{\Delta}=0$.
For the sake of definiteness, $\omega=10^{-2}\Lambda_k$. In addition, $\tilde{\Delta}(\omega)=\Delta(\omega)/\Lambda_k$ and $\tilde{V}(\omega)=V(\omega)/|V_{\rm crit}|$, where $V_{\rm crit}$ is given in Eq.~(\ref{App-Delta-0-crit-2D}). In addition, we used a finite frequency cutoff $\Lambda_{\omega}=2\Lambda_k$ in the numerical calculations.
}
\label{fig:App-Gap-Eq-inter-SS-swave-Delta-2D}
\end{figure*}

The electron DOS is defined in Eq.~(\ref{App-Observables-DOS-def}). For the even- and odd-frequency pairings it reads as
\begin{eqnarray}
\label{App-Observables-DOS-even-2D}
\nu^{\rm even}(\omega) &=& \frac{\sign{\omega}}{2\pi} \int_0^{\Lambda_k/v_F} \frac{dk}{v_F \mu} \left\{(\omega-\mu)\left[(\omega+\mu)^2 - |\Delta(\omega)|^2 -v_F^2k^2\right] -2\mu |\Delta(\omega)|^2\right\} \theta(\omega^2-|\Delta(\omega)|^2) \nonumber\\
&\times&\left\{\delta\left[\omega^2 - |\Delta(\omega)|^2 -\left(\mu +v_Fk\right)^2\right] -\delta\left[\omega^2 - |\Delta(\omega)|^2 -\left(\mu -v_Fk\right)^2\right] \right\}
\end{eqnarray}
and
\begin{eqnarray}
\label{App-Observables-DOS-odd-2D}
\nu^{\rm odd}(\omega) &=& \frac{\sign{\omega}}{2\pi} \int_0^{\Lambda_k/v_F} \frac{dk}{v_F \sqrt{|\Delta(\omega)|^2+\mu^2}} \left\{(\omega-\mu)\left[(\omega+\mu)^2 - |\Delta(\omega)|^2 -v_F^2k^2\right] -2\mu |\Delta(\omega)|^2\right\} \nonumber\\
&\times&\left\{\delta\left[\omega^2 -\left(v_Fk+\sqrt{\mu^2 +|\Delta(\omega)|^2}\right)^2\right] -\delta\left[\omega^2 -\left(v_Fk-\sqrt{\mu^2 +|\Delta(\omega)|^2}\right)^2\right] \right\},
\end{eqnarray}
respectively. The results for the electron DOS at $\Delta(\omega)=\alpha$ and $\Delta(\omega)=\alpha \Lambda_k/\omega$ as well as $\Delta=\alpha \omega/\sqrt{\omega^2+\beta^2\Lambda_k^2}$ and $\Delta=\alpha \omega \Lambda_k/(\omega^2+\beta^2\Lambda_k^2)$ are presented in Figs.~\ref{fig:App-Observables-DOS-even-odd-2D} and \ref{fig:App-Observables-DOS-4-2-2D}, respectively. As expected, the DOS in 2D is linear in $\omega$ and shows the same cusp-like features as in the 3D case (cf. with results in Sec.~V). The spectral functions are identical, therefore, we do not present them here.

\begin{figure*}[!ht]
\begin{center}
\includegraphics[height=0.4\textwidth]{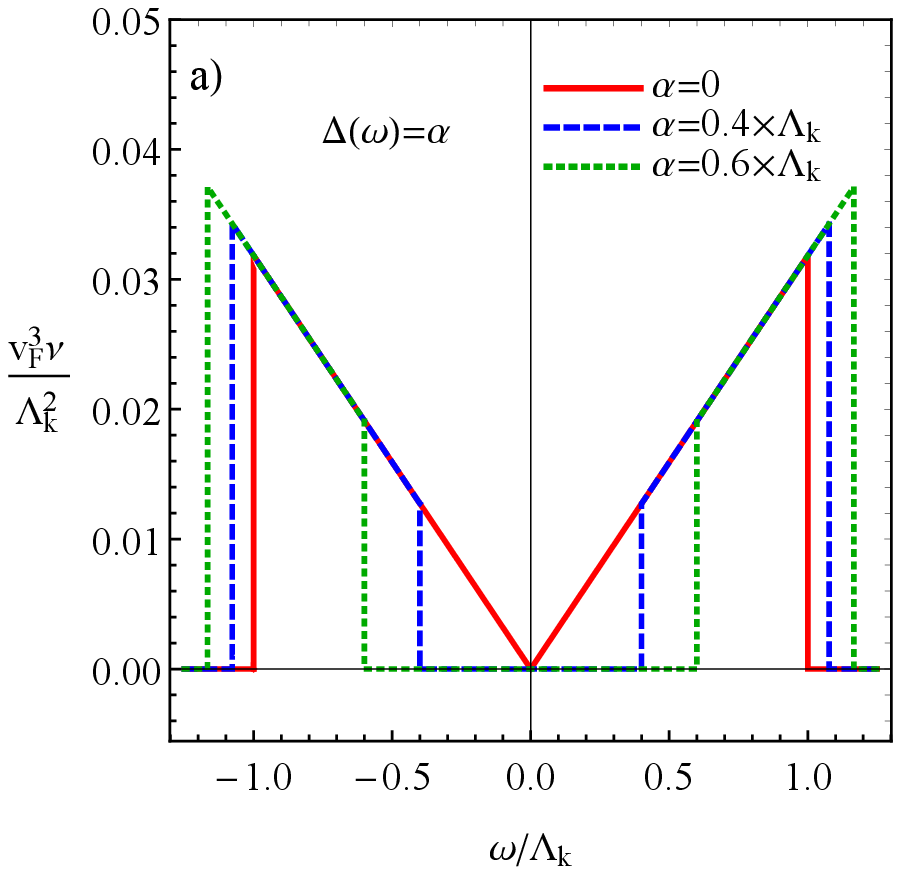}\hfill
\includegraphics[height=0.4\textwidth]{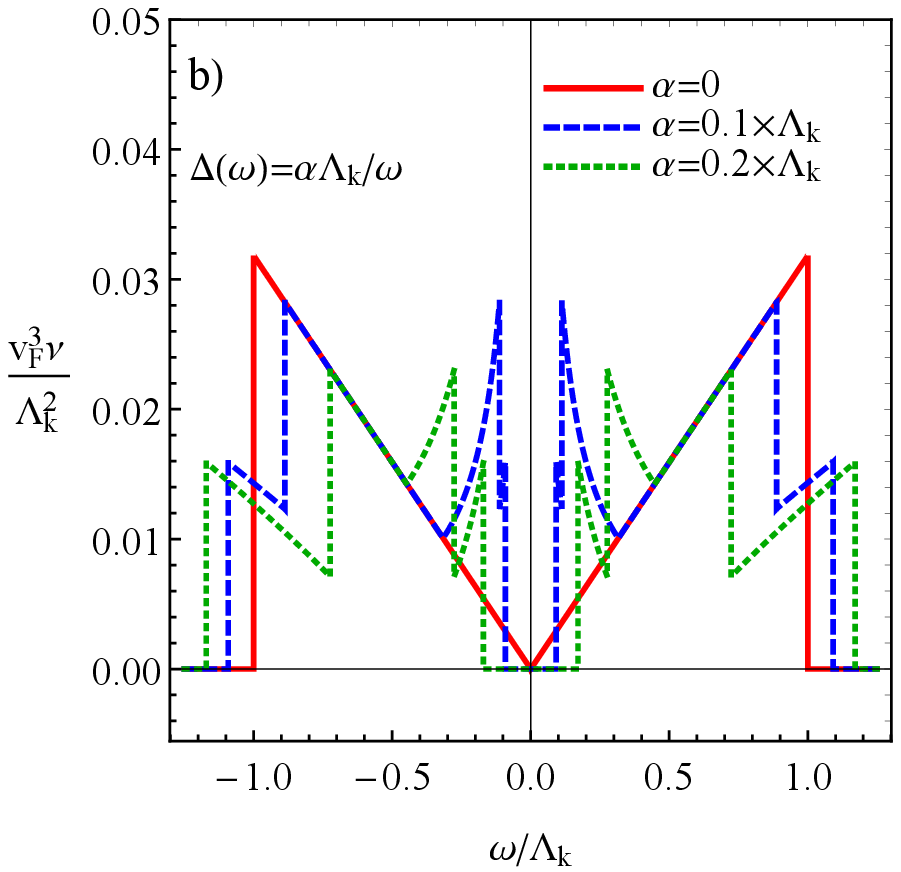}
\end{center}
\caption{The dependence of the electron DOS $\nu(\omega)$ on the frequency $\omega$ for $\Delta(\omega)=\alpha$ (panel a)) and $\Delta(\omega)=\alpha \Lambda_k/\omega$ (panel b)) at a few values of the gap strength $\alpha$. We set $\mu=0$.
}
\label{fig:App-Observables-DOS-even-odd-2D}
\end{figure*}

\begin{figure*}[!ht]
\begin{center}
\includegraphics[height=0.4\textwidth]{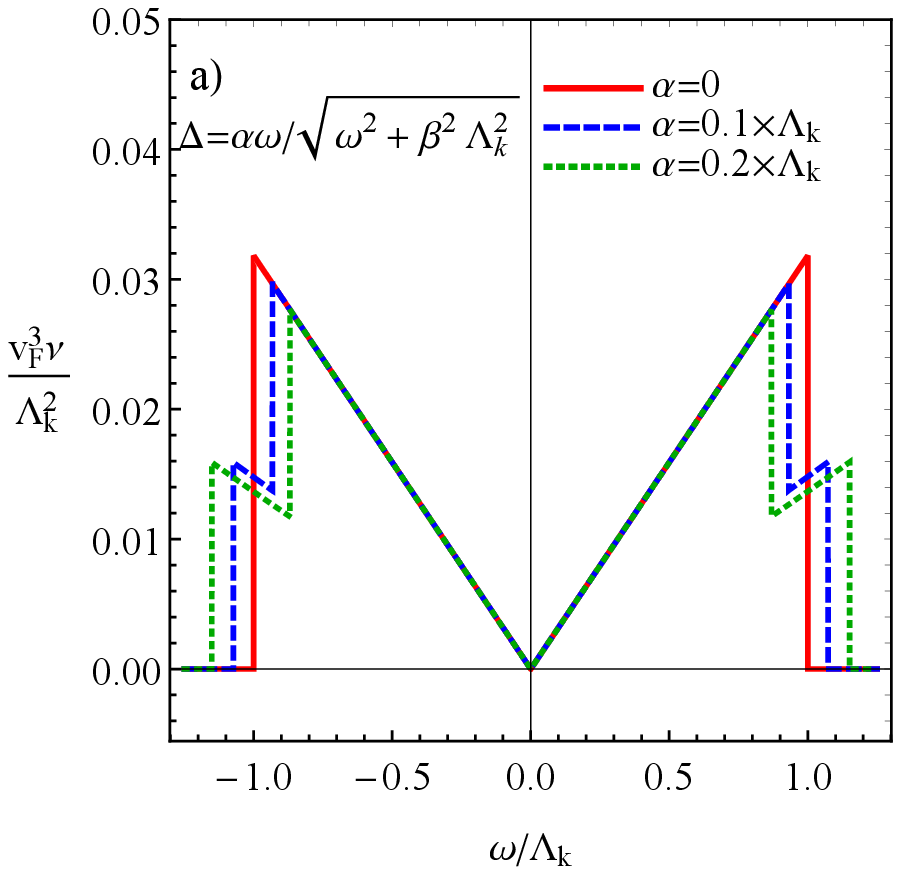}\hfill
\includegraphics[height=0.4\textwidth]{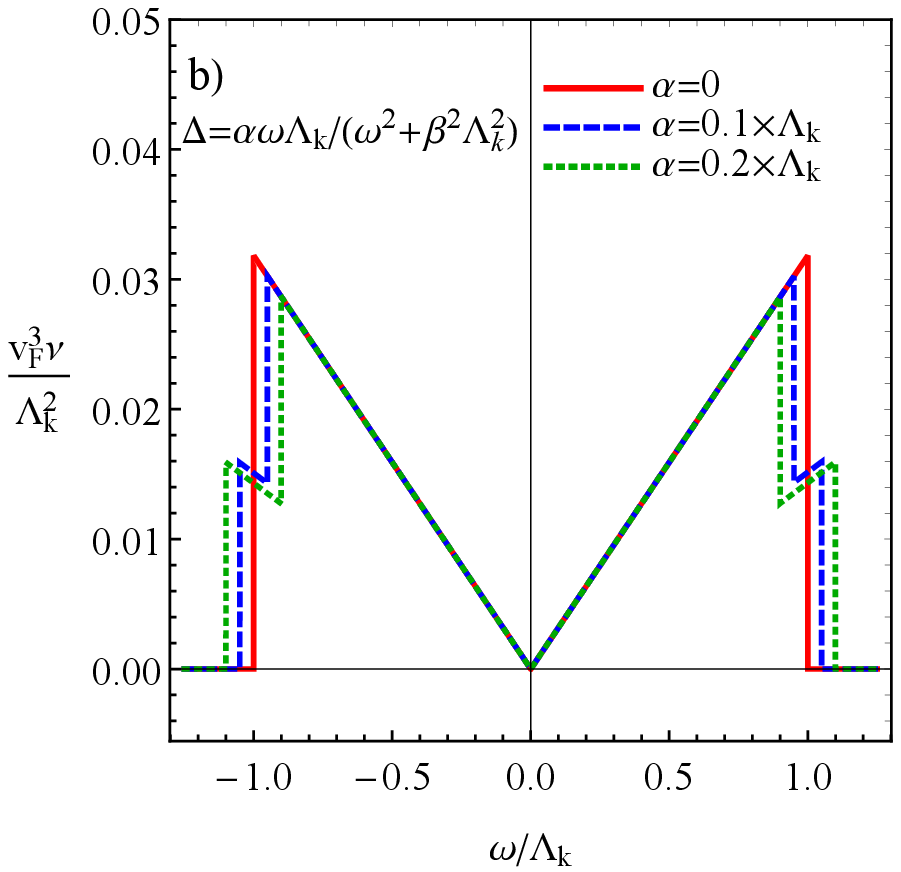}
\end{center}
\caption{The dependence of the electron DOS $\nu(\omega)$ on the frequency $\omega$ for $\Delta=\alpha \omega/\sqrt{\omega^2+\beta^2\Lambda_k^2}$ (panel a)) and $\Delta=\alpha \omega \Lambda_k/(\omega^2+\beta^2\Lambda_k^2)$ (panel b)) at a few values of the gap strength $\alpha$. We set $\mu=0$ and $\beta=1$.
}
\label{fig:App-Observables-DOS-4-2-2D}
\end{figure*}

Finally, we compare the dependence of the electron DOS $\nu(\omega)$ on $\omega$ for $\Delta=\alpha$ and $\Delta=\alpha \Lambda_k/\omega$ at nonzero $\mu$. The corresponding results are presented in Figs.~\ref{fig:App-Observables-DOS-mu-2D}a) and \ref{fig:App-Observables-DOS-mu-2D}b), respectively. Like in 3D case, while the coherence peaks appear for $\Delta=\alpha$ at $\mu\neq0$, they are absent for $\Delta=\alpha \Lambda_k/\omega$.

\begin{figure*}[!ht]
\begin{center}
\includegraphics[height=0.4\textwidth]{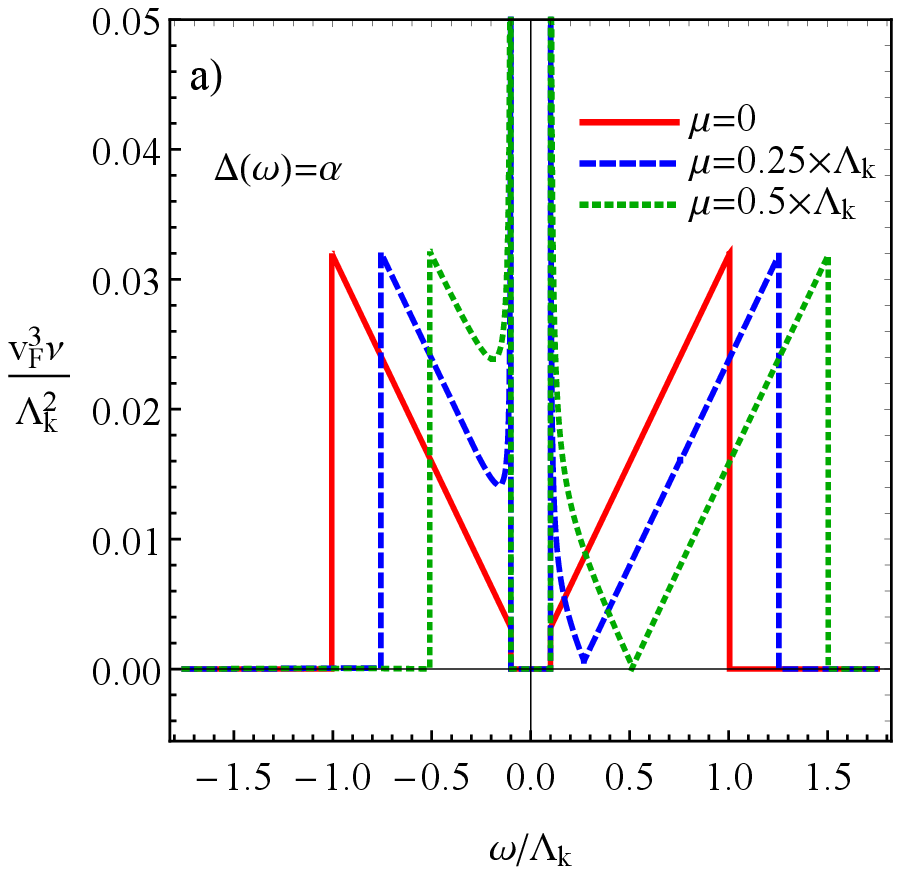}\hfill
\includegraphics[height=0.4\textwidth]{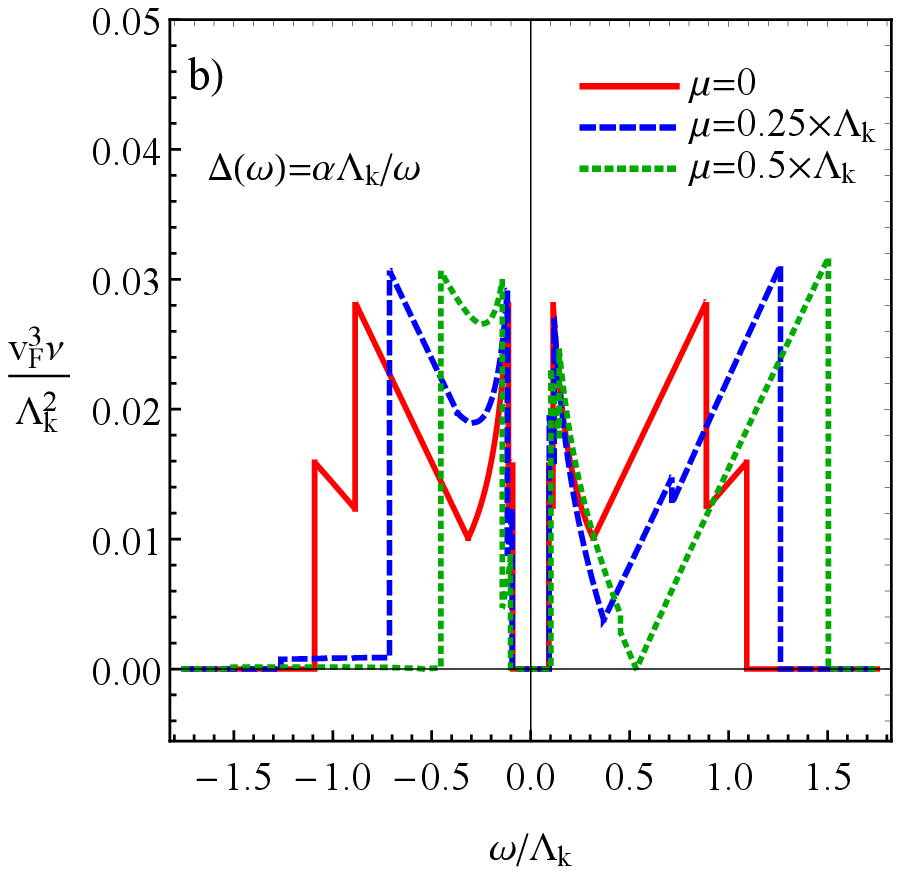}
\end{center}
\caption{The dependence of the electron DOS $\nu(\omega)$ on the frequency $\omega$ for $\Delta=\alpha$ (panel a)) and $\Delta=\alpha \Lambda_k/\omega$ (panel b)) at a few values of the electric chemical potential $\mu$. We set $\alpha=0.1\,\Lambda_k$.
}
\label{fig:App-Observables-DOS-mu-2D}
\end{figure*}

\vspace{0.5cm}
\begin{center}
\noindent\rule{6cm}{1pt}
\end{center}
\vspace{0.5cm}

\begin{itemize}

\item[[S1\!\!]] 
    V.~L.~Berezinskii,
    Pis'ma Zh. Eksp. Teor. Fiz {\bf 20}, 628 (1974).

\item[[S2\!\!]] 
    J.~Linder and A.~V.~Balatsky,
    arXiv:1709.03986.

\item[[S3\!\!]] 
    D.~Solenov, I.~Martin, and D.~Mozyrsky,
    Phys. Rev. B {\bf 79}, 132502 (2009).

\item[[S4\!\!]]	
    H.~Kusunose, Y.~Fuseya, and K.~Miyake,
    J. Phys. Soc. Jpn. {\bf 80}, 054702 (2011). 

\item[[S5\!\!]] 
    A.~Altland and B.~D.~Simons, {\sl Condensed Matter Field Theory} (Cambridge University Press, 2010).

\item[[S6\!\!]]
    J.~W.~Negele and H.~Orland, {\sl Quantum Many-Particle Systems} (CRC Press, 1988).

\item[[S7\!\!]]
    T.~O.~Wehling, A.~M.~Black-Schaffer, and A.~V.~Balatsky,
    Adv. Phys. {\bf 63}, 1 (2014).

\item[[S8\!\!]] 
    G.~E.~Volovik, {\sl The Universe in a Helium Droplet} (Oxford University Press, 2003).

\item[[S9\!\!]]
    Z.~Wang, X.-L.~Qi, and S.-C.~Zhang,
    Phys. Rev. Lett. {\bf 105}, 256803 (2010).

\end{itemize}
\end{widetext}

\end{document}